\documentclass[%
 reprint,
superscriptaddress,
 amsmath,amssymb,
 aps,
prb,
]{revtex4-2}

\usepackage{graphicx}
\usepackage{dcolumn}
\usepackage{bm}
\usepackage{braket}
\usepackage[]{graphicx}
\usepackage[dvipsnames]{xcolor}
\usepackage{hyperref}
\usepackage[anythingbreaks]{breakurl}

\hypersetup{
    colorlinks=true,
    linkcolor=blue,
    filecolor=blue,      
    urlcolor=blue,
    citecolor=blue,
    breaklinks=true
}

\usepackage{siunitx} 

\newcommand{\lzprobability}[0]{P_{\text{LZ}}}
\newcommand{\amp}[0]{A}
\newcommand{\gradbx}[0]{\Delta B_x}
\newcommand{\gradbz}[0]{\Delta B_z}
\newcommand{\tunnelcoupling}[0]{t_c}
\newcommand{\detuning}[0]{\epsilon}
\newcommand{\tanhRange}[0]{R_\text{tanh}}
\newcommand{\zeemansplitting}[0]{E_z}
\newcommand{\Tres}{T_{\text{res}}}
\newcommand{\omres}{\omega_{\text{res}}}
\newcommand{\cdc}{c_{\text{dc}}}
\newcommand{\timestrech}{s_t}
\newcommand{\frabi}{\omega_R}
\newcommand{\trace}{\text{tr}}

\begin{document}

\preprint{APS/123-QED}

\title{Flopping-Mode Electron Dipole Spin Resonance in the Strong-Driving Regime}

\author{Julian D. Teske}
\email{julian.teske@rwth-aachen.de}
\author{Friederike Butt}
\author{Pascal Cerfontaine}
\affiliation{
JARA-FIT Institute for Quantum Information, Forschungszentrum Jülich GmbH and RWTH Aachen University,
52074 Aachen, Germany
}

\author{Guido Burkard}
\affiliation{
Department of Physics, University of Konstanz, 
78457 Konstanz, Germany
}%

\author{Hendrik Bluhm}%
\affiliation{
JARA-FIT Institute for Quantum Information, Forschungszentrum Jülich GmbH and RWTH Aachen University,
52074 Aachen, Germany
}

\date{\today}
             
\begin{abstract}
Achieving high fidelity control of spin qubits with conventional electron dipole spin resonance (EDSR) requires large magnetic field gradients of about $\SI{1}{\milli \tesla \per \nano \meter}$, which also couple the qubit to charge noise, and large drive amplitudes of order 1 mV. 
The flopping-mode is an alternative method to drive EDSR of an electron in a double quantum dot, where the large displacement between both dots increases the driving efficiency. 
We propose to operate the flopping-mode in the strong-driving regime to use the full magnetic field difference between the two dots.
In simulations, the reduced required magnetic field gradients suppress the infidelity contribution of charge noise by more than two orders of magnitude, while providing Rabi frequencies of up to $\SI{60}{\mega \hertz}$.
However, the near degeneracy of the conduction band in silicon introduces a valley degree of freedom that can degrade the performance of the strong-driving mode. This necessitates a valley-dependent pulse optimization and makes operation to the strong-driving regime questionable.

\end{abstract}

\maketitle


\section{Introduction}
In recent years, semiconductor qubits have developed tremendously with key results being fidelities above the error correction threshold \cite{Noiri2022HighFidGates, Xue2022HighFidGates} and small-scale quantum processors \cite{6qubitPhilips2022, 2qubitprocessorHighfid2022}, see \cite{Burkard2021} for a recent review. New proposals to construct a universal quantum computer based on this qubit type have been formulated \cite{BlueprintShuttle, SpiderWebArray}. Supporting the long term perspective, semiconductor qubits were shown to be compatible with classical CMOS-technology \cite{Xue2021CMOSCryo, Pauka2021CMOSControlChip} and can be operated at temperatures above $\SI{1}{\kelvin}$ \cite{Petit2020UniversalHotQubits, Yang2020processor1kelvin}. Furthermore, there has been progress in the transport of single charges \cite{Mills2019} and of spin qubits through semiconductor quantum dot arrays, with the objective of establishing medium-range coupling interaction \cite{ConveyorInga, Shuttling1dot}, offering much higher connectivity than next-nearest neighbor coupling.

A well established technique for the manipulation of a single spin in a quantum dot (QD) is the electron dipole spin resonance (EDSR), where a spin orbit coupling can be naturally present in the material \cite{EDSRSpinOrbit} or artificially induced by an inhomogeneous magnetic field \cite{EDSRMicroMagnet}. This spin-orbit coupling is used to drive spin transitions by resonantly shifting the position of the QD  with an electric signal. Another method is the driving of electron spin resonance (ESR) with microwaves emitted by a nanoscale transmission line in proximity of the QDs. However, EDSR is often considered as the method of choice because it allows driving with purely electric fields, which are easier to localize and dissipate less energy than the AC-currents needed for ESR. The energy dissipation is especially important when cryogenic electronics are involved in the pulse generation.

To identify possibilities to improve EDSR, a detailed assessment of its requirements and limitations is required.
While fidelities of $99.95 \%$, sufficient for quantum error correction, have been reached, 
the required strong magnetic field gradients couple the electron spin to electric noise, which was identified as the main limitation of the achievable fidelity \cite{Yoneda2018HighFidGates, Struck2020}.
In conclusion, a method requiring smaller field gradients would let the qubit tolerate larger electric noise, for example due to a higher operating temperature or noisy, power limited cryoelectronics.

Furthermore, the strong confinement on an electron in a QD necessitates relatively strong driving signals with a substantial power dissipation to reach typical Rabi frequencies in the $\si{\mega \hertz}$ range \cite{Yoneda2018HighFidGates}. 
Working at lower magnetic fields would be advantageous because the resulting lower drive frequencies can be expected to reduce crosstalk and simplify the realization of ultra-low-power cryoelectronic control solutions. 
However, the total magnetic field $\vec{B}_0$ must be compatible with the required magnetic field gradient, as the local inhomogeneous magnetic fields $\vec{B}_{\text{loc}}(\vec{x})$ are usually provided by micro-magnets, which are magnetized by $\vec{B}_0$ \cite{NeumannMicroMagnet}.
Large field gradients require optimized magnet designs with special shapes, which increase the complexity of designing scalable layouts. Moreover, 
it is difficult to achieve a total magnetic field  $\vec{B}(\vec{x}) = \vec{B}_{\text{loc}}(\vec{x}) + \vec{B}_0$ much smaller than the micromagnet field, as the externally applied field $\vec{B}_0$ would then have to largely cancel out the micromagnet field. Foreseeable variation of the latter between qubits will hamper a strong cancelation for many qubits at a time. 

One approach to working with lower magnetic field gradients is the flopping-mode driving scheme, which applies the EDSR principle to an electron confined in a double quantum dot (DQD). Instead of shifting the position of a single QD, the driving electric signal changes the potential difference of the two dots in the DQD, displacing the electron within the DQD. This displacement provides a large electric dipole moment, while being more efficient in terms of driving power than shifting a QD \cite{crootflopping, benitoflopping}.
The distance of the two dots is usually on the order of $\SI{100}{\nano \meter}$, whereas the displacement of the QD in conventional EDSR is typically only $\SI{1}{\nano \meter}$. This increases the dipole moment and allows the use of much smaller magnetic field gradients and smaller total magnetic fields. Smaller Zeeman splittings are also beneficial for spin relaxation time, especially in the case of silicon, if the Zeeman splitting is smaller than the valley splitting \cite{HuSpinHotspot}. 

The flopping-mode qubit first attracted attention, because its strong electric dipole moment enables strong spin-photon coupling in cavities \cite{Input-outputBenito, Mi2018FloppingSpinPhotonInterface, FloppingSpinPhotonCouplingSi, CouplingSingleElectronMW, StrongCouplingfloppingCavity}. This spin-photon coupling can also be used to mediate two-qubit gates \cite{Flopping2QubitCavityMediated}. Alternatively, two-qubit gates for the flopping-mode can be implemented by capacitive coupling \cite{flopping2qubitcapacitively} and the flopping-mode can even be applied to heavy-hole-qubits in Germanium \cite{FloppingGeHole}.

While the weak driving regime considered in previous studies is most efficient when aiming for the lowest possible drive amplitude \cite{crootflopping}, improving the fidelity favors large drive amplitudes, thus eventually leaving the weak driving regime.
We thus propose and analyse a strong-driving regime for the flopping-mode qubit with a pulse amplitude much larger than the tunnel coupling of the DQD such that the electron is entirely shifted from one dot to the other, exploiting the full magnetic field gradient across the DQD. 
The central idea of our proposal is to let the electron oscillate resonantly between the stable and noise insensitive positions of strong confinement in the left and right QD while spending as little time in the transition between the dots. 

The concept of moving the electron between the discrete positions in the left and the right QD can also be supported by the driving pulse. We study a smoothed rectangular driving pulse that provides the maximal time of strong confinement between adiabatic transitions. As the electron is strongly confined in one QD most of the time, small perturbation to the pulse have only a weak effect on the electrons position. 

Parallel to our studies, the strongly confined configuration was considered for the robust storage for quantum information and the transition into the weak-driving flopping mode regime for manipulation was investigated \cite{FloppingModeRegimeTransition}. This idea switches the strong dipole moment on and off between storage and manipulation to improve the noise robustness during idle time, while our proposal intends to use the robustness of strong confinement to improve the robustness of the manipulation.

The strong-driving flopping-mode qubit suits proposals for sparse architectures for a semiconductor spin quantum processor \cite{SpiderWebArray}, where DQDs are readily available without requiring additional sites and cross-talk is suppressed by the large distances between qubits. 
The low magnetic field is also beneficial in shuttling-based quantum processors \cite{BlueprintShuttle}. Furthermore, the proposed pulse scheme poses only minimal requirements concerning the pulse generation electronics.

We simulate the flopping mode to calculate the achievable fidelity in presence of noise. We show the reduced noise sensitivity by the transition from the weak- to the strong-driving regime. In the strong-driving regime we discuss two different pulse shapes to fully leverage the potential of flopping mode.
We identify optimal parameter regions where a good noise resilience is achieved while the probability for orbital excitation is minimized. Leakage by orbital excitation poses the limiting factor for the performance in the general model (without valley states). We calculate excellent fidelities with an improvement of more than two orders of magnitude in fidelity compared to conventional EDSR.

To discuss the realization of the strong-driving regime of the flopping mode in Si-based devices, we include different valley states in our model. The valley splitting has been argued to be a local stochastic material parameter \cite{Gefluctuations} that is difficult to control. Valley excitations can degrade the performance of the flopping mode so that pulses have to be optimized specifically for a given valley splitting. In the strong-driving regime, high-fidelity manipulation can only be realized with a favorable valley splitting configuration so that high-fidelity operation may in general be constrained to the weak-driving regime. 

The remainder of this article is structured as follows. In Sec.~\ref{sec::model}, we present our qubit model and two pulse schemes to drive the flopping-mode qubit. In Sec.~\ref{sec::results}, we present the results of our simulation. In Sec.~\ref{sec::valley}, we extend the simulation to discuss implications of the valley degree of freedom (DOF) for flopping-mode qubits in silicon and in Sec.~\ref{sec::sumandoutlook}, we give an outlook and summarize our results.

\section{Model}
\label{sec::model}

\begin{figure}
    \centering
    \includegraphics{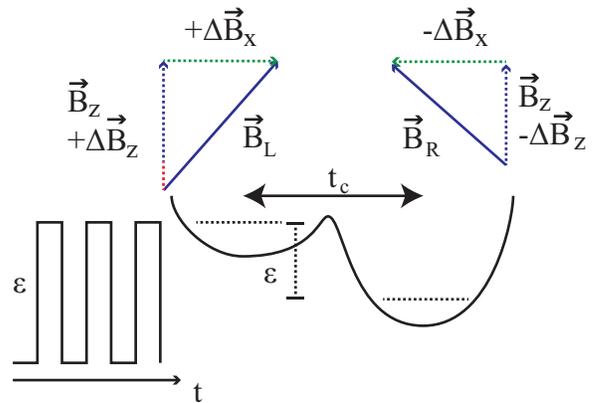}
    \caption{Sketch of the flopping-mode qubit. An electron is trapped in a DQD of tunnel coupling $\tunnelcoupling$ and the detuning $\detuning$ corresponds to the energy difference of the ground state in the left and right QD. The electron's wave function is shifted between the left and right QD by an electric pulse on the detuning. Here, a rectangular pulse is drawn for the manipulation. The magnetic field gradient $\Delta B_x$ introduces an artificial spin-orbit coupling that allows Rabi-driving of the spin state. The total magnetic field $\vec{B}_z$ determines the Zeemann energy and hence the resonance frequency of the qubit.}
    \label{fig:skech}
\end{figure}

\begin{figure}
    \centering
    \includegraphics{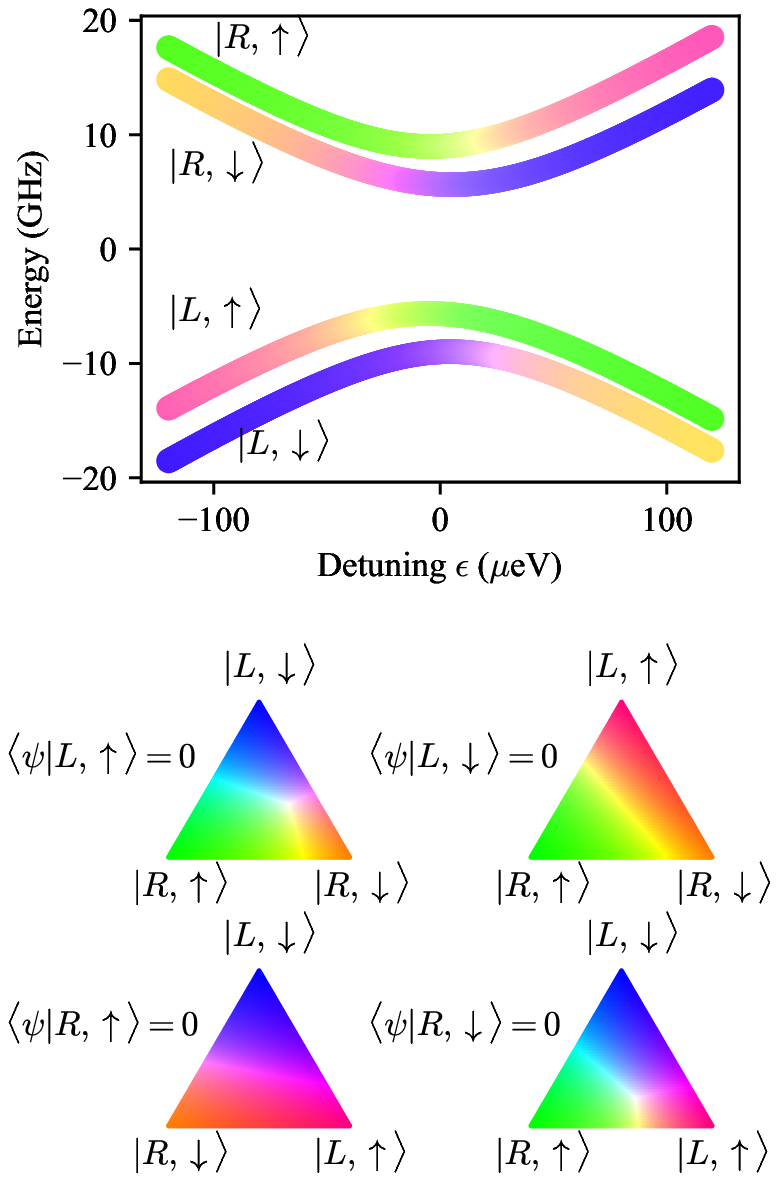}
    \caption{Energy Spectrum of the flopping-mode qubit. The energies of the eigenstates are plotted against the detuning of the DQD. The colors encode the changes in the eigenstates as visualized by the triangles below. Each triangle shows the color code for superpositions of three basis states. Each corner of the triangles corresponds to one basis state, each edge to superpositions of two basis states and the interior of the triangle is filled with non-zero superpositions of all three eigenstates. The orbital states show an avoided crossing due to the tunnel coupling $\tunnelcoupling=\SI{30}{\micro \electronvolt}$. The energy of the spin states is split by the Zeeman Energy $\zeemansplitting / g\mu_B=\SI{120}{\milli \tesla}$ with the asymmetry introduced by the magnetic gradient along the z-axis $\Delta B_z = \SI{40}{\milli \tesla}$. The two lowest lying energy states differ by their spin and transitions can be driven by the coupling magnetic gradient $\Delta B_x = \SI{60}{\milli \tesla}$. The large magnetic field values were chosen to increase the clarity of the illustration.}
    \label{fig:EnergySpectrumSpinOrbit}
\end{figure}

\begin{figure*}
    \centering
    \includegraphics{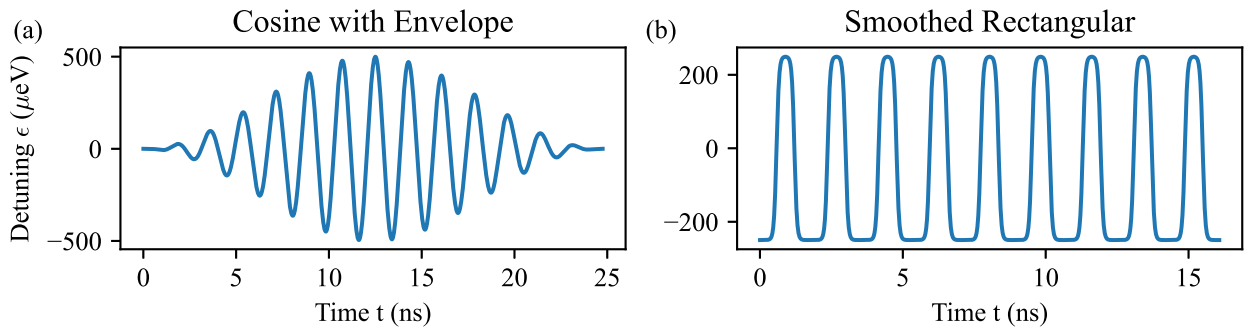}
    \caption{Plots of the two considered pulse forms.
    (a) The cosine pulse with a sinusoidal envelope serves as reference pulse form. The pulse starts and ends in a symmetrically tuned DQD.
    (b) The smoothed rectangular pulse with a steepness parameter of $\tanhRange = 13$ represents the proposed pulse form. The asymmetry between positive and negative values is set by the duty cycle parameter $\cdc=0.3$. The rectangular pulse starts and ends in a strongly detuned DQD such that the electron is confine in one QD. Both pulses were plotted for illustration with $A=\SI{250}{\micro \electronvolt}$, $\Tres = 2 \pi / \omega \approx \SI{.56}{\nano \second}$, and $T\approx \SI{25}{\nano \second}$ and $T\approx \SI{16}{\nano \second}$ for the cosine and rectangular pulse respectively.}
    \label{fig:pulses}
\end{figure*}

We model the flopping-mode qubit as single electron in a DQD placed in an inhomogeneous magnetic field as sketched in Fig.~\ref{fig:skech}. We label the QDs left (L) and right (R) and assume different magnetic field values in the QDs $\vec{B}_L$ and $\vec{B}_R$. The coordinate system is chosen such that the total magnetic field $\vec{B}_L + \vec{B}_R$ is aligned with the z-axis and the perpendicular component $\Delta\vec{B}_x$ of magnetic field gradient is aligned with the x-axis.  We publish an implementation of the qubit model and the pulse shapes in the qopt-applications repository \cite{qopt-applications}.

\subsection{Hamiltonian}
The Hamiltonian of an electron in a DQD with the convention $\hbar=1$ is given by \cite{benitoflopping}
\begin{align} \label{eq:hamiltonian}
\begin{split}
    H(\detuning(t)) = & \frac{\detuning(t)}{2}\tau_z + t_c \tau_x + \frac{\zeemansplitting}{2} \sigma_z \\
    &+ \frac{g \mu_B }{2} \left( \gradbx \sigma_x  + \gradbz \sigma_z  \right) \tau_z,
\end{split}
\end{align}
where the detuning $\detuning$ is the difference in the electric potentials of the two dots as function of time $t$ and the tunnel coupling $\tunnelcoupling$ of the two dots is assumed constant. The Zeeman energy is denoted $\zeemansplitting = g \mu_B B_z$ with the effective g-factor $g$, Bohr magneton $\mu_B$, and the magnetic field gradient in x-direction (z-direction) $\gradbx$ ($\gradbz$). The Pauli matrices on the spin and orbital DOF are denoted $\sigma$ and $\tau$, and operate on the basis $(\ket{\uparrow} , \ket{\downarrow} )$  and $(\ket{L}, \ket{R})$ respectively. The orbital degree of freedom $(\ket{L}, \ket{R})$ is not to be confused with different orbital states of an electron in a single dot. Shifts of the QDs (used for traditional EDSR) are not included. These shifts contribute to variations of the magnetic fields, but can be neglected, because the magnetic field gradients are about 100 times smaller than those used for standard EDSR \cite{Yoneda2018HighFidGates} in the regime of interest.

The energy spectrum of the Hamiltonian defined in Eq.~\eqref{eq:hamiltonian} is plotted in Fig.~\ref{fig:EnergySpectrumSpinOrbit}. The assumption of a small Zeeman splitting compared to the tunnel coupling $\zeemansplitting \ll \tunnelcoupling$ leads to a dominant orbital splitting and a large avoided crossing between eigenstates with different orbital states. Thus, with our manipulation method we want to drive spin excitations between the ground state and the first excited state while being adiabatic with respect to the orbital DOF. This driving is mediated by $\gradbx$ while we assume a finite $\gradbz$ to account for a placement tolerance of the micro-magnet. 
For concreteness, we assume a conservative ratio of $\gradbx / \gradbz = 5$, whereas even higher ratios have already been realized \cite{crootflopping}. 

We include three noise contributions in our model and we use the term \textit{quasistatic} for noise contributions at frequencies much slower than the qubit dynamics. First, we consider noise on the detuning $\detuning \rightarrow \detuning(t) + \delta \detuning(t)$, where quasistatic noise $\delta \detuning$ is sampled from a Gaussian distribution. The second contribution is described as fast white noise $\delta \detuning(t)$ with a constant spectral noise density $S$ up to a finite cutoff frequency. Third, we describe the hyperfine interaction as quasistatic magnetic noise on the Zeeman splitting $\zeemansplitting \rightarrow \zeemansplitting + g \mu_B \delta B_z$ with a standard deviation $\sigma_B $. The contribution of charge noise to the quasistatic magnetic noise is suppressed by the low magnetic field gradients.

\subsection{Pulse Shapes}

We investigate two pulse shapes. The first one is a cosine pulse with an envelope and the second pulse is a smoothed rectangular function (Fig.~\ref{fig:compareCosineRect}(a)). The cosine pulse is given by
\begin{align} \label{eq:cosine_pulse}
    \epsilon(t) = \epsilon_0 + 2 \sin^2 \left(\frac{\pi t}{T} \right) A \cos(\omega t + \phi),
\end{align}
with the offset $\detuning_0$, amplitude $A$,  total pulse time $T$,  carrier frequency $\omega$ and  phase offset $\phi$. The $\sin^2$ term acts as an envelope and is scaled by the factor of two to have unity average amplitude. 

The cosine pulse is not ideal concerning the three key properties heat dissipation,  achievable Rabi frequency and noise susceptibility. In terms of heat dissipation, in the strong driving regime $\amp \gg 2 \tunnelcoupling$, the occupation probability in one dot saturates such that the highest detuning peaks of the sine function do not contribute to the displacement of the electron, while they still contribute to the heat emission of the high frequency signal. The achievable Rabi frequency is limited because the electron is not fully displaced during the initial and final oscillations of the pulse due to the envelope. And finally the noise susceptibility is not ideal yet, because the electron's wave function is stretched over the DQD for a large portion of the pulse yielding a high spin-orbit mixing on average.

We envision the ideal pulse to meet the following three requirements to improve with respect to the three key properties. First, the pulse starts and terminates localized in one QD. Second, the electron is strongly localized in one QD during as much of the pulse duration as possible without the amplitude reaching unnecessarily high values. Third, the electron is shifted between the two dots as fast as possible to reduce noise susceptibility, while still being transferred adiabatically with regard to both charge and spin excitations.

We implement such a pulse with a smoothed rectangular pulse shape  (Fig.~\ref{fig:compareCosineRect}(b)). The rectangular pulse is modelled in the simulation by transitions shaped as hyperbolic tangent functions.
\begin{align} \label{eq:rectangular_pulse}
\begin{split}
        \detuning(t) & = \amp \tanh{\left( \frac{ 2\tanhRange}{\Tres} \left(t-  \frac{\Tres(1 + \cdc )}{4} \right)\right)}, \\
        & \forall t: t = n \Tres + s , s < \Tres /2, n \in \mathbb{N}, \\
    \detuning(t) & = -\amp   \tanh{\left(\frac{2\tanhRange }{\Tres}\left(t-  \frac{\Tres(3 - \cdc)}{4} \right)\right)}, \\
    & \forall t: t = n \Tres + s , s > \Tres /2, n \in \mathbb{N},
\end{split}
\end{align}
where $\Tres = \frac{2 \pi}{\zeemansplitting}$ is the resonance oscillation period, $\tanhRange$ the pulse steepness and the duty cycle parameter $\cdc \in [0, 1]$ effectively controls the Rabi frequency. This control is required, because the amplitude cannot be used for the control of the Rabi frequency in the strong-driving regime due to saturation effects. The duty cycle parameter introduces an asymmetry such that the electron remains longer in one QD than in the other as can be seen in Fig.~\ref{fig:pulses}(b). Intuitively, $\cdc$ determines the time the spin is exposed to the driving field. For $\cdc=0$ the driving is symmetric between both QDs providing the largest Rabi frequency possible and for $\cdc=1$ the electron remains stationary in one QD. Negative values for $\cdc$ are possible but have the same effect with exchanged QDs and are not used in our formulation. The pulse steepness $\tanhRange$ is introduced to control the velocity of the charge transfer within the DQD, where a large $\tanhRange$ corresponds to a sharper rectangle and therefore a faster charge transfer.

We require the total pulse time to be an integer multiple of the resonance period time, such that the electron is always driven in full cycles and terminates in the same QD as it started. This complicates the pulse tuning as it makes the time a discrete optimization parameter. We use the duty cycle parameter to adjust the effective driving strength to the pulse time. Note that two single-qubit gates about orthogonal axes can be acquired by a phase shift of the pulse by $\pm\pi/4$.

\subsection{Methods}

We performed quantum dynamics simulations of the proposed pulses using the simulation and optimal control package qopt \cite{qoptPaper}. Before the optimal performance of the flopping-mode qubit can be evaluated, the pulses must be optimized for a given magnetic field and DQD parameters, because the inaccuracy of analytic estimates of the optimal pulse parameters would deteriorate the qubit performance. The numeric optimization allows us to estimate the optimal pulse length and frequency $(T, \omega)$ for the cosine pulse or the optimal pulse length and duty cycle parameter $(T, \cdc)$ for the rectangular pulse with great accuracy.

A suitable figure of merit for the quantification of the qubit performance needs to account for the spin dynamics and excitations in the orbital DOF, which we consider to be leakage. Therefore, we define an infidelity $\mathcal{I}_L$ in the presence of leakage between the calculated propagator $U$ and a target unitary $U_{\text{target}}$ as
\begin{align} \label{eq:infid}
    \mathcal{I}_L(U_{\text{target}}, U) = 1 - \frac{1}{4}|\trace(U_{\text{target}}^\dagger U\vert_{V_C} )|^2,
\end{align}
where $U\vert_{V_C}$ denotes the truncation of the propagator to the computational subspace.
The measure $\mathcal{I}_L$ always considers leakage as erroneous and thus $\mathcal{I}_L>0$ for a finite probability of orbital excitation during the pulse. For example, if we take the target to be identity $U_{\text{target}} = I$ and the propagator to include orbital excitation $U(x)=\exp(xi\pi \tau_x / 2)$, then we calculate $\mathcal{I}_L(I, U(x))=1 - (\cos(x \pi) + 1) / 2$.
The definition is inspired by common fidelity measures \cite{LeakageCostFunction}.
Our optimization target $U_{\text{target}}$ is an $X_\pi$-gate in the computational space, but the pulse-optimization algorithm can be adapted to any rotation around the x-axis. 
Other fidelity measures can also be used to asses the performance in specific situations. For example, if the fidelity of an experiment shall be predicted, then the simulation can be extended by an orbit-independent spin measurement, implemented as partial trace over the orbital DOF. The average gate fidelity for such a quantum channel can be calculated following \cite{LeakageCostFunction, PRBHolonomicAverageGateFid}. Although the exact fidelities are deviating, we found the systematic relations and the interpretation to be independent of the fidelity measure.

We define the computational space in terms of the eigenvectors $v_i, 1 \leq i \leq 4$ of the Hamiltonian at $t=0$. We want to drive excitations in the spin DOF, so we define the computational space $V_C$ to be the vector space that is spanned by the ground and the first excited state with a different spin. With the assumption that $\zeemansplitting < 2 \tunnelcoupling$, we have:
\begin{align}
    V_C = \text{span}(v_1, v_2).
\end{align}
If the orbital splitting $\Omega = \sqrt{\detuning^2 +4\tunnelcoupling^2}$ set by the tunnel coupling and offset in detuning becomes smaller than the Zeeman splitting, we need to replace $v_2$ with $v_3$. The leakage space is then defined to be spanned by the remaining eigenvectors. These definitions are justified by the assumption that the flopping-mode qubit is initialized and read out adiabatically. For the initialization, this means that the qubit is initialized as a Loss-DiVincenzo qubit and then adiabatically transferred into the DQD with detuning $\detuning(t=0)$. To perform the read out, the electron is first adiabatically confined in one QD and then read out as Loss-DiVinvenzo qubit.

To study the noise resilience of the proposed driving mode, we perform Monte Carlo noise simulations of quasistatic and white noise. For the simulation of quasistatic noise, we average over 8 noise values $\delta \detuning$ to approximate an integral over a Gaussian distribution. For white noise we generate 1000 time-dependent noise traces $\delta \detuning(t)$ from the spectral noise density with pseudo random numbers. 
The effect of quasistatic noise is calculated in a basis adiabatically following the noise, meaning that the eigenvectors defining the computational space are calculated for the Hamiltonian $H(\detuning(t=0) + \delta \detuning)$.

The tuning procedure is described in more detail in the supplementary material in 
Sec.~\ref{app::optimizaiton}.
We verified the simulation by reproducing previous experimental and theoretical results as discussed in Appendices \ref{app::croot} and \ref{app::benito}.

\section{Results}
\label{sec::results}

\begin{figure*}
    \centering
    \includegraphics[]{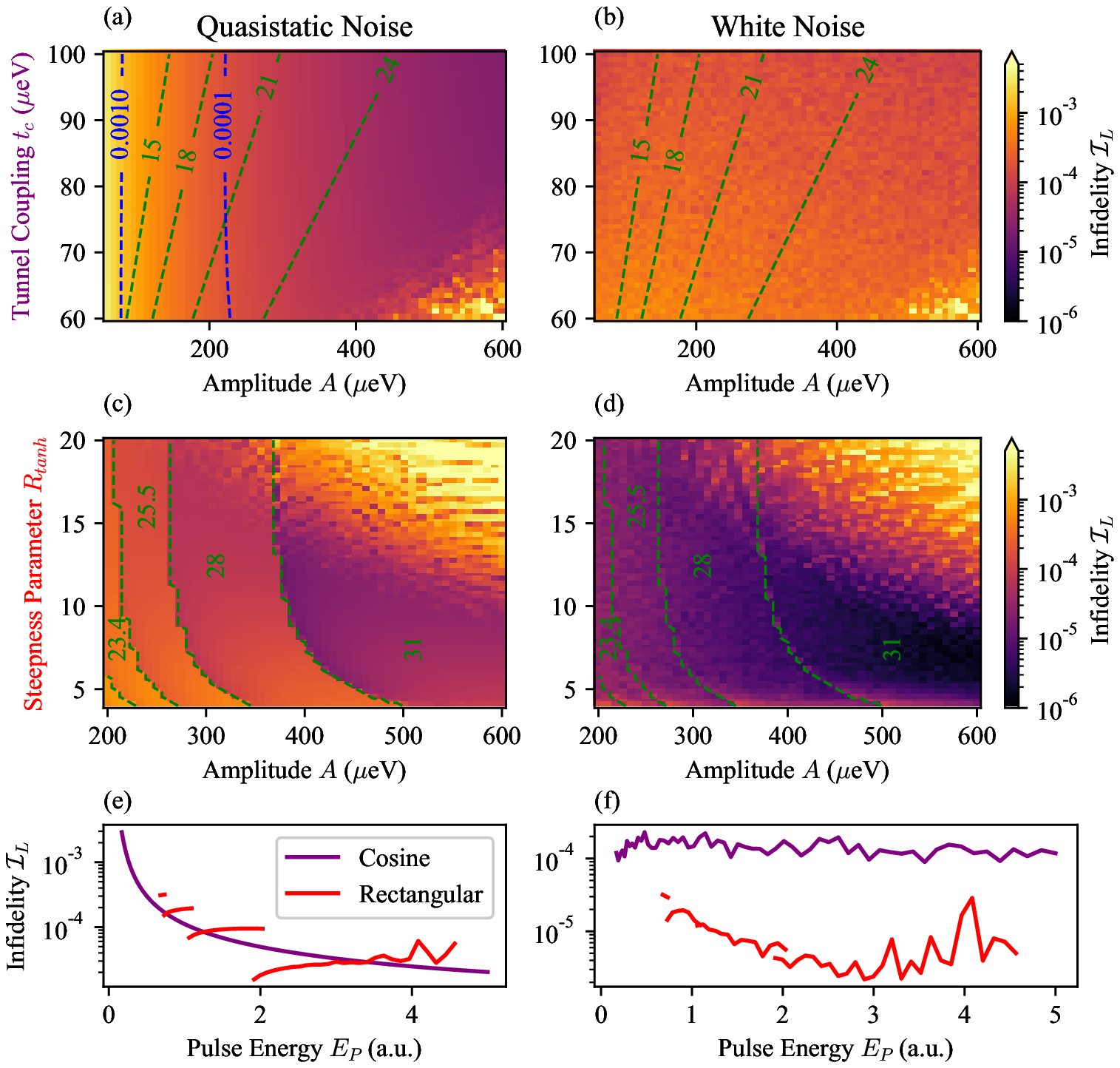}
    \caption{Infidelity of the $X_\pi$ gate calculated with Monte Carlo simulations of a flopping-mode qubit driven by a cosine pulse (a, b) or a smoothed rectangular pulse (c, d). The simulations include quasistatic noise on the control signal with a standard deviation of $\sigma_\detuning =\SI{15}{\micro \electronvolt}$ in (a, c, e) and white noise with a spectral noise density of $\sqrt{S} = \SI{0.07}{\nano \electronvolt \per \sqrt \hertz}$ in (b, d, f).
    Both simulations are performed with magnetic fields of $\Delta E_z / (g \mu_B)= \SI{20}{\milli \tesla}$, $\Delta B_x = \SI{2}{\milli \tesla}$, $\Delta B_z= \SI{0.4}{\milli \tesla}$. The simulation of the rectangular pulse features a large tunnel coupling of $\tunnelcoupling=\SI{100}{\micro \electronvolt}$. The green dotted lines in (a, c) mark the contour lines of the Rabi frequency with values in $\si{\mega \hertz}$.
    (a) Contour lines of the fidelity are plotted in blue. The infidelity introduced by quasistatic noise is reduced for higher amplitudes unless LZ excitations increase the leakage for large amplitudes and small tunnel couplings.
    (b) The infidelity contribution by white noise is largely independent of the amplitude and tunnel coupling, unless LZ excitations are dominant.
    (c) We can identify an optimal region around $A=\SI{400}{\micro \electronvolt}$ and $\tanhRange=12$. For small amplitudes the pulse becomes slower and more susceptible to quasistatic noise while LZ excitations are dominant for steep pulses with a large amplitude. The Rabi frequency changes substantially at the discontinuities of the fidelity and is otherwise almost constant. The green Rabi frequency values correspond to the plateaus.
    (d) The rectangular pulse is robust towards white noise in a wide range of parameters. 
    The plots in (e) and (f) show cuts along (a, c) and (b, d) respectively. The data is cut along $\tunnelcoupling=\SI{100}{\micro \electronvolt}$ and $\tanhRange=9.97$ to consider data of the same tunnel coupling and the optimal steepness $\tanhRange$. Instead of the amplitude $A$, we plot the infidelity against the corresponding power dissipation $E_P(A)$. The rectangular pulse can achieve a higher fidelity and larger Rabi frequency with less power dissipation.
    } 
    \label{fig:compareCosineRect}
\end{figure*}

\begin{figure*}
    \centering
    \includegraphics[]{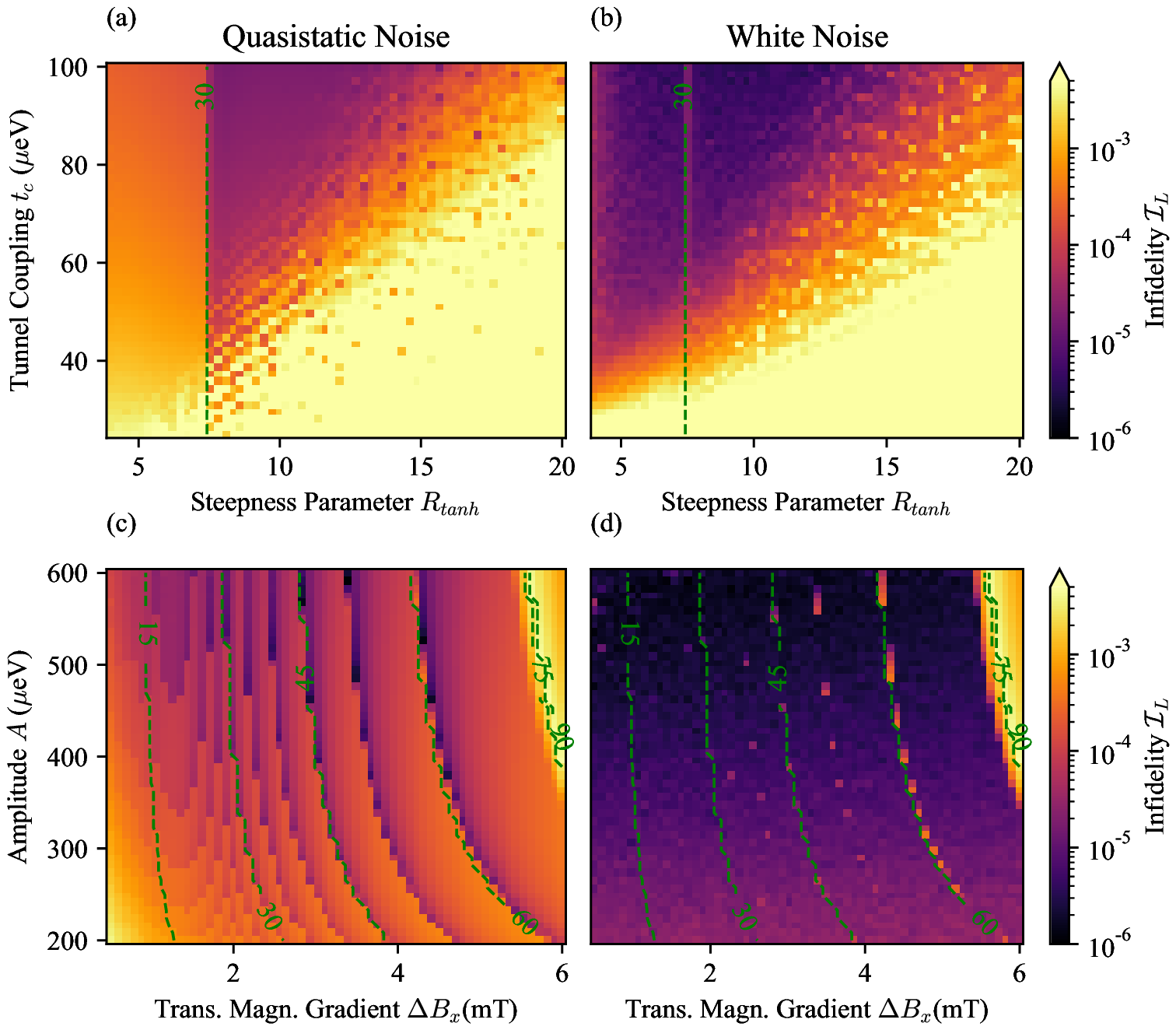}
    \caption{Infidelity of the $X_\pi$ gate calculated with Monte Carlo simulations of a flopping-mode qubit driven by a smoothed rectangular pulse. The simulations include quasistatic noise on the control signal with a standard deviation of $\sigma_\detuning =\SI{15}{\micro \electronvolt}$ in (a, c) and white noise with a spectral noise density of $\sqrt{S} = \SI{0.07}{\nano \electronvolt \per \sqrt \hertz}$ in (b, d).
    Both simulations were performed with magnetic fields of $\Delta E_z / (g \mu_B)= \SI{20}{\milli \tesla}$ and $\Delta B_z= \SI{0.4}{\milli \tesla}$. The simulation in (a, b) features a transversal magnetic gradient of $\Delta B_x = \SI{2}{\milli \tesla}$ and a constant relation of amplitude to the tunnel coupling $A=4\tunnelcoupling$
    and the simulation in (c, d) features a large tunnel coupling of $\tunnelcoupling=\SI{100}{\micro \electronvolt}$ and a steepness parameters of $\tanhRange=8$.
    All gate infidelities were averaged over the simulated noise traces as Monte Carlo simulation. The green dotted lines in (a, c) mark the contour lines of the Rabi frequency.
    (a) The plot is split by the contour line at about $\tanhRange=7.5$ in two parts, because at this value, the pulse is extended by an additional period of the resonance frequency. In the left part of the plot, the susceptibility to quasi static noise is increased by the larger asymmetry of the pulse. 
    (b) The white noise couples to LZ excitations and is otherwise suppressed. The white noise is independent of the discreteness of the pulse.
    (c) There is a large range of parameters, where the pulse is protected from quasi static noise. We can clearly see the discreteness effects of the pulse. The driving scheme breaks down at extremely high transversal magnetic gradients around $\Delta B_x = \SI{6}{\milli \tesla}$, as the Rabi frequency approaches the resonance frequency.
    (d) The rectangular pulse is robust towards white noise in a wide range of amplitudes and magnetic gradients.
    } 
    \label{fig:LZSandFastPulse}
\end{figure*}

In this section, we consider  sources of strong quasistatic electric noise with a standard deviation of $\sigma_\detuning=\SI{15}{\micro \electronvolt}$ and white noise with a spectral noise density of $S=\SI{0.07}{\micro \electronvolt \per \sqrt \hertz}$ and a cutoff at $\SI{10}{\giga \hertz}$. We chose noise values larger than those typically measured in experiments \cite{Yoneda2018HighFidGates, DialNoiseSpectroscopy} to demonstrate the noise resilience of the proposed driving mode.
We choose to evaluate the flopping-mode qubit in a small magnetic field $E_z / (g \mu_B) = \SI{20}{\milli \tesla}$ with a transversal magnetic field gradient $B_x = \SI{2}{\milli \tesla}$ and a longitudinal gradient $B_z = \SI{0.4}{\milli \tesla}$, unless stated otherwise. 

We also consider quasistatic fluctuations of the Zeeman splitting $\zeemansplitting$ with a standard deviation of $\sigma_B =\SI{3}{\micro \tesla}$. Such a magnetic noise can for example originate from fluctuations of nuclear spins that couple to the electron spin in the DQD. Our value for $\sigma_B$ is chosen lower than measured values in natural silicon \cite{HFINaturalSiKawakami2014-ek} but higher than measurements in isotopically purified silicon suggest \cite{Struck2020}. The value we chose corresponds to a residual $^{29}$Si concentration of $0.5\%$ or a dephasing time of $T^\ast_2=\hbar/\sigma_B\approx \SI{1.9}{\micro \second}$ \cite{ConcentrationSi29HyperfineTheoretical}. Even for the relatively high $\sigma_B$, perturbations from hyperfine noise can be neglected as they are dominated by the electrical noise and leads to an additional infidelity of about $10^{-5}$, depending mainly on the pulse time $T$.
We verified the consistency of our simulations with previous theoretical (see Appendix~\ref{app::benito}) and experimental results (see 
Appendix~\ref{app::croot}).

\subsection{Pulse Shape Comparison}

A direct comparison of the cosine and the rectangular pulse is made in Fig.~\ref{fig:compareCosineRect}. The graphs (a) and (b) show the performance of the cosine pulse in the presence of electric noise. In (a), a clear improvement in the fidelity with larger amplitude can be observed except for very large amplitude values paired with a small tunnel coupling. In this case the infidelity is dominated by leakage into the excited orbital state caused by diabatic Landau-Zener (LZ) excitation. The green dotted lines are contour lines of the Rabi frequency showing the acceleration of the pulse with rising amplitude. The blue lines are contour lines of the infidelity. Note that the contour lines for the infidelity and the Rabi frequency are not parallel. This indicates that the increase in fidelity is not determined by the relation of the dephasing time to the pulse time $T$ alone. Instead, it originates partially from the stronger confinement in one of the dots by larger absolute pulse values, since the system is less susceptible to noise when the detuning $\detuning$ is large and the spin-orbit mixing low.

Panels (c) and (d) in Fig.~\ref{fig:compareCosineRect} show corresponding plots for the rectangular pulse, where $\tunnelcoupling$ is kept constant, because $\tunnelcoupling = \SI{100}{\micro \electronvolt}$ is the global optimum (compare Fig.~\ref{fig:LZSandFastPulse}) and $\tanhRange$ is varied to map out the strong-driving regime. The step-like features in the plot (c) are caused by the discreteness in the tuning of the rectangular pulse. The rectangular pulse time $T$ must always be a multiple of the resonance time $T=n\frac{2 \pi}{E_z} ; n \in \mathbb{N}$. At every change of $T$, the Rabi frequency is increased and also the duty cycle parameter $\cdc$ has a discontinuity. On each step, the fidelity drops and then continuously increases with the amplitude. This behavior is linked to $\cdc$, because large duty cycle values correspond to asymmetric pulses, which are more susceptible to electric noise. The largest infidelity values in (c) and (d) arise for very steep pulses at large amplitudes and this infidelity contribution is caused by LZ excitation. The peaks in plot (d) (more clearly visible in Fig.~\ref{fig:LZSandFastPulse}~(d)) near the green lines reflect coherent errors that are caused by pulse tuning. At these values, the total pulse time is shorter than ideal and the duty cycle can only decrease the Rabi frequency. The optimum for the fidelity is reached at a discontinuity, which is a pathological case, because the exact position of the discontinuity and the values in its proximity depend on the choice of the pulse length. Changes to the optimization algorithm can vary the results at these points (compare Appendix~\ref{app::optimizaiton}).
However, this is not a serious limitation to the driving mode, because the fidelity only degrades of about a factor of two when choosing pulse parameters away from the discontinuity.

Even though both pulses can reach robust noise insensitivity, the rectangular pulse has an advantage in terms of power dissipation. To quantify the pulse energy we define
\begin{align}
    E_P \propto \int (\detuning(t) - \detuning_0)^2 dt,
\end{align}
and plot in panes (e) and (f) the infidelity $\mathcal{I}_L$ against $E_P$ for horizontal cuts through the plots (a)-(d). (e) shows that a smaller infidelity can be reached with the rectangular pulse with a given $E_P$, if an optimal duty cycle can be chosen. From a comparison of the contour lines in (a) and (c) we can also extract that the rectangular pulse provides higher Rabi frequencies. The comparison of (b) and (d) also demonstrates that the rectangular pulse is less sensitive towards white noise in a wide region.

\subsection{Landau-Zener Excitation}

Having identified leakage as one of the major sources of infidelity, we investigate the influence of leakage more closely. The shift of the electron from one dot to another can be theoretically described as a LZ transition. LZ theory predicts an excitation probability of $\lzprobability = \exp(-2\pi \delta)$ with $\delta = \tunnelcoupling^2 / 4v$ with the level velocity $v=\frac{d \detuning}{d t}$ \cite{LZS_SHEVCHENKO20101}. A tight upper bound for the velocity of the rectangular pulse described in \eqref{eq:rectangular_pulse} can be calculated with a first order Taylor expansion of the hyperbolic tangent to
\begin{align}
    v \leq \frac{2 \amp \tanhRange}{\Tres} = \frac{\amp \zeemansplitting \tanhRange}{\pi}.
\end{align}
Thus, operation with low $\zeemansplitting$ is beneficial for the suppression of LZ excitation in addition to the advantages discussed above. In case of the amplitude $A$, a trade-off must be made be since lower amplitudes further suppress LZ excitation and reduce the dissipated energy but also increase the noise sensitivity. The plot in Fig.~\ref{fig:LZSandFastPulse} (a, b) explores the range of possible values for $\tunnelcoupling$ and $\tanhRange$, while keeping the relation $A=4\tunnelcoupling$ to remain in the strong driving regime. We observe that a higher tunnel-coupling is generally favorable, while there is a sweet spot for $\tanhRange$ between an increase in LZ excitation and an increased noise sensitivity. In the absence of noise, we can clearly observe Landau-Zener-Stückelberg oscillations as discussed in the 
Appendix~\ref{app::LZS}.

The discreteness of the rectangular pulse can also be observed in Fig.~\ref{fig:LZSandFastPulse}. Like in Fig.~\ref{fig:compareCosineRect} (c, d), discontinuities in the sensitivity to quasistatic noise appear when the pulse time $T$ changes. We do not observe an asymmetry around the steps for the infidelity caused by white noise. In Fig.~\ref{fig:LZSandFastPulse} (b, d) there are only tuning artifacts along the changes in the Rabi frequency, where the total pulse time is not optimal.

\subsection{High Rabi Frequencies}

The achievable Rabi frequency is mainly determined by the transversal magnetic field gradient $\Delta B_x$, while the influence of $A$ saturates for large $A$ values. The infidelity as function of $\Delta B_x$ and $A$ is plotted in Fig.~\ref{fig:LZSandFastPulse} (c, d). The Rabi frequency $\omega_R$ can be brought up to values of more than $\SI{60}{\mega \hertz}$, indicating that the rotating wave approximation $\frac{\zeemansplitting}{\omega_R}\gg1$ does not need to be fulfilled. The increased Rabi frequency reduces the influence of magnetic noise and for the assumed standard deviation of $\sigma_B = \SI{3}{\micro \tesla}$ the infidelity contribution drops to $10^{-6}$.

\subsection{Comparison to conventional EDSR}

The noise robustness of the flopping-mode qubit becomes evident in the direct comparison with EDSR in a single QD. In this section, we estimate the infidelity that arises, when a single QD is driven by EDSR and exposed to a charge noise source of the strength considered as the quasistatic electric noise above.
We start with the rotating frame Hamiltonian for an electron in a QD under resonant drive, given by
\begin{align} \label{ham:driven_qubit}
    H = \frac{\delta \omega}{2}  \sigma_z  + \frac{\omega_R}{2} \sigma_x,
\end{align}
where $\delta \omega$ denotes the frequency detuning of the driving signal from the resonance frequency and $\omega_R$ denotes the Rabi frequency. High fidelities were achieved by driving the electron with an electric control signal in a local magnetic field gradient with a transversal $b_{\text{trans}}$ and a longitudinal $b_{\text{long}}$ gradient. The transversal gradient determines the Rabi frequency $\omega_R = A_r g \mu_B b_{\text{trans}} / 2$ with the amplitude of the displacement $A_r$.

The predominant contribution to the frequency detuning $\delta \omega$ originates from charge noise, which displaces the QD by a distance of $\delta r$ and couples to the longitudinal magnetic field gradient $\delta \omega = \delta r g \mu_B b_{\text{long}}$ \cite{Yoneda2018HighFidGates}.
Both the driving pulse and the noise fluctuations shift the position of the QD over a distance of $\delta r$ by creating a local electric field of strength $\delta E$. When approximating a QD as an harmonic confining potential with an orbital energy of $ \Delta_{\text{orb}}$, then the relation between positional shifts and applied electric fields is given by:
\begin{align}
    \delta r = \frac{\hbar^2 e}{m_{\text{Si}} \Delta_{\text{orb}}} \delta E,
\end{align}
where $m_{\text{Si}}$ denotes the effective electron mass in silicon and $e$ the elementary charge (see supplementary material of \cite{Yoneda2018HighFidGates}).
Further, the electric field can be described by a shift in the electric potential $\delta V$:
\begin{align}
    \delta E = \frac{\delta V}{e d},
\end{align}
where the distance $d$ is measured from the QD to the charge defect creating $\delta E$. Let's assume that any defects causing charge noise and the driving metallic gates have the same distance to the QD of approximately $d\approx \SI{100}{\nano \meter}$.

The infidelity caused as pure dephasing by quasistatic noise for an $X_\pi$-gate up to leading order is given by
\begin{align}
    \mathcal{I}\approx \left(\frac{\langle\delta \omega\rangle}{\omega_R}\right)^2 = \left(\frac{2 \langle \delta V\rangle }{A } \frac{b_{\text{long}}} {b_{\text{trans}}}\right)^2.
\end{align}

A previous experiment achieved an optimal fidelity at a Rabi frequency of $\omega_R = \SI{3.9}{\mega \hertz}$, corresponding to a driving amplitude of $A=\SI{70}{\micro \electronvolt}$ for the magnetic field gradients $b_{\text{trans}}= \SI{1}{\milli \tesla \per \nano \meter} $ and $b_{\text{long}}= \SI{0.2}{\milli \tesla \per \nano \meter}$ \cite{Yoneda2018HighFidGates}. We can now extrapolate the performance of this experiment for our assumption of quasi-static electric noise with a standard deviation of $\langle \delta V \rangle=\SI{15}{\micro \electronvolt}$ to an infidelity of $\mathcal{I}=0.7\%$, which is more than two orders of magnitude larger compared to the optimal values in the flopping mode, where an infidelity of about $\mathcal{I}_L \approx 10^{-5}$ is predicted. In conclusion, the direct comparison to EDSR in a single QD for our assumed noise values demonstrates the much lower noise sensitivity of the flopping-mode.

\section{Flopping-Mode EDSR in Silicon} 
\label{sec::valley}

\begin{figure*}
    \centering
    \includegraphics[]{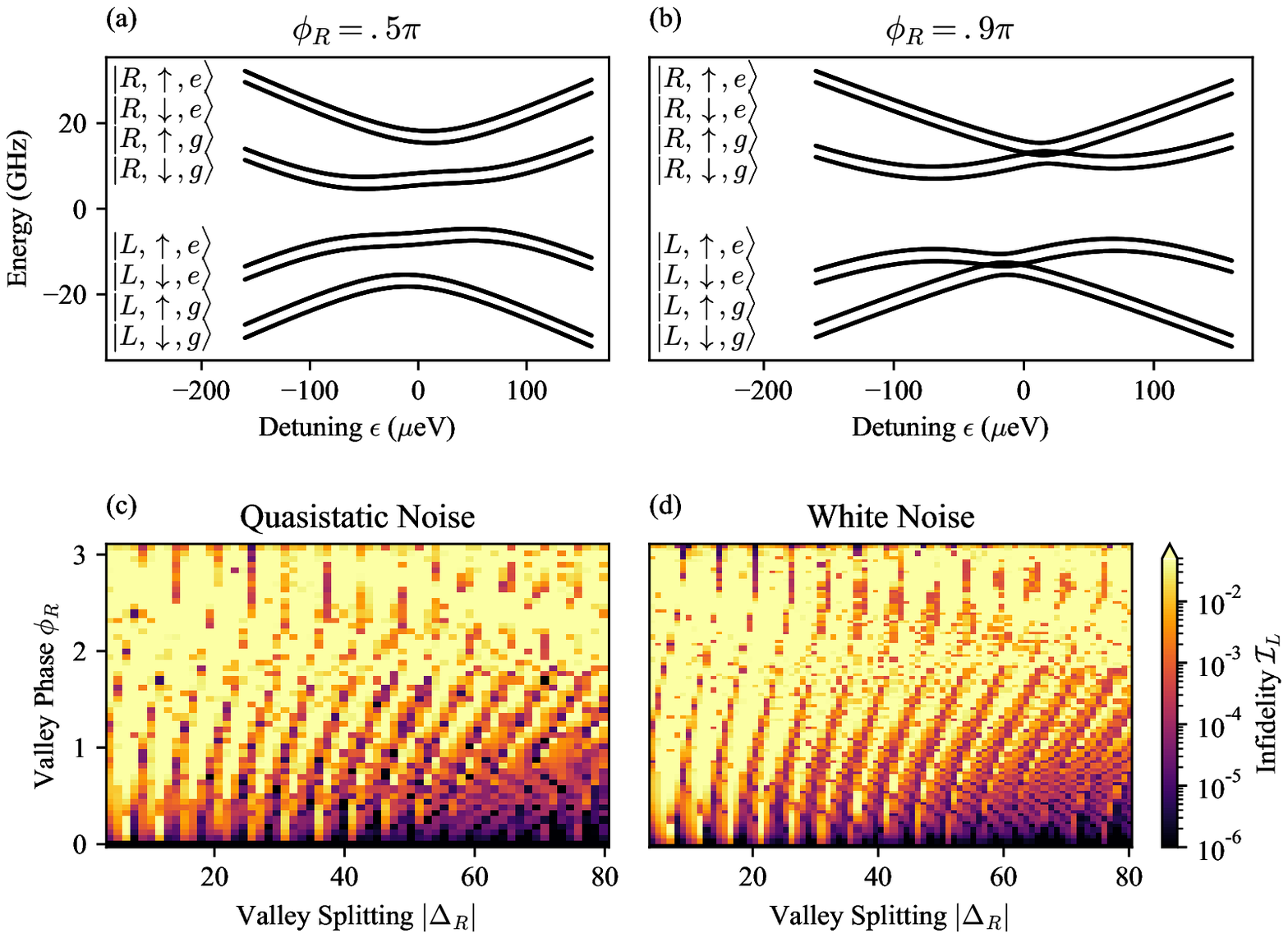}
    \caption{
    Simulation of the valley model.
    (a, b) show the energy spectra of a DQD with the additional valley DOF. The spectra are plotted for $\tunnelcoupling =\SI{40}{\micro \electronvolt}$, $\Delta E_z / (g \mu_B)= \SI{100}{\milli \tesla}$ (chosen for clarity), $\Delta B_x = \SI{20}{\milli \tesla}$, $\Delta B_z= \SI{10}{\milli \tesla}$, $|\Delta_R| = \SI{40}{\micro \electronvolt}$. 
    (a) is calculated for a valley phase of $\phi_L = .5 \pi$ and (b) for a valley phase of  $\phi_L = .9 \pi$, where additional avoided crossings appear.
    (c, d) show Monte Carlo simulations of a flopping-mode qubit driven by a smoothed rectangular pulse including the valley DOF. The simulation was performed with magnetic fields of $\Delta E_z / (g \mu_B)= \SI{20}{\milli \tesla}$, $\Delta B_x = \SI{2}{\milli \tesla}$, $\Delta B_z= \SI{0.4}{\milli \tesla}$ and the parameters $\tunnelcoupling=\SI{100}{\micro \electronvolt}$, $A \approx \SI{422}{\micro \electronvolt}$, $\tanhRange \approx 8.1$, and $\Delta_L = \SI{60}{\micro \electronvolt}$.
    (c) shows a Monte Carlo simulation of quasistatic noise with a standard deviation of $\sigma_\detuning =\SI{15}{\micro \electronvolt}$. The infidelity shows strong oscillations in the valley phase. If the pulse can be tuned away from a maximum of the oscillation, then the valley splitting should allow high-fidelity gates.
    (d) shows a Monte Carlo simulation of white noise with a spectral noise density of $\sqrt{S} = \SI{0.07}{\nano \electronvolt \per \sqrt \hertz}$. The infidelity shows the same oscillations in the valley phase as in (c). 
    } 
    \label{fig:valley}
\end{figure*}

\begin{figure}
    \centering
    \includegraphics[]{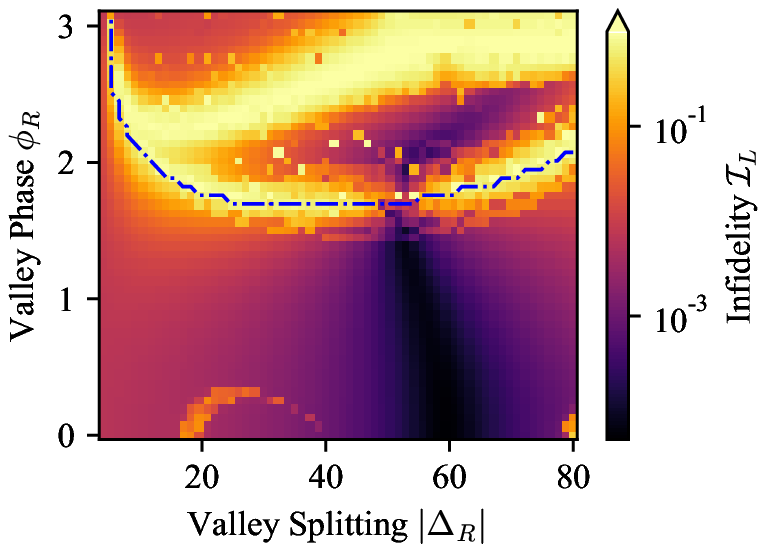}
    \caption{
    Influence of the valley splitting in the weak driving regime. The simulation was performed with experimentally realized parameters \cite{crootflopping} with a magnetic field with $\Delta E_z / (g \mu_B)= \SI{209}{\milli \tesla}$, $\Delta B_x = \SI{15}{\milli \tesla}$ and  $\Delta B_z= \SI{0.27}{\milli \tesla}$, and a tunnel coupling of $\tunnelcoupling=\SI{23}{\micro \electronvolt}$. We assumed $\Delta_L = \SI{60}{\micro \electronvolt}$ just like in the simulation for Fig.~\ref{fig:valley} and simulated a pulse of cosine shape as in Fig.~\ref{fig:pulses}(a) with an amplitude of $A=\SI{0.42}{\micro \electronvolt}$ and an offset $\detuning_0=\SI{5.4}{\micro \electronvolt}$. Additionally we simulated quasi static electric noise with a standard deviation of $\sigma_\detuning =\SI{0.5}{\micro \electronvolt}$. Along the blue dash-dotted line, the first two excited states are degenerate. Below the line, we drive transitions to the first excited state and above the line to the second excited state.
    } 
    \label{fig:weak_drive_valley}
\end{figure}

In this section, we discuss the prospects of realizing flopping-mode qubits in silicon, as it is the currently predominant material for the construction of spin qubit quantum processors \cite{6qubitPhilips2022, 2qubitprocessorHighfid2022}.
The most relevant peculiarity of silicon is the conduction band degeneracy leading to the presence of a valley DOF. Silicon hetero-structures posses two low lying valley states that can influence the spin dynamics. Usually, the influence is modeled by a valley-dependent g-factor or a valley-orbit coupling mechanism. 

The electron band structure of a two-dimensional electron gas in silicon has a two-fold degenerate minimum \cite{ValleyBasicsAndo1982}. In the heterostructures used for the fabrication of semiconductor spin qubits, this degeneracy is lifted by the sharp potential step at a silicon-insulator interface \cite{ValleyInterfacePhysics, ValleySplittingEffectiveMass}. The energy difference between these two valley states is called the valley splitting.
We extend the Hamiltonian of \eqref{eq:hamiltonian} by the valley DOF
\begin{align} \label{eq:HamiltonianValley}
    H_V = H  + \sum_{i=L,R} P_i 
    \begin{pmatrix}
    0 & \Delta_i \\
    \Delta_i^\ast & 0
    \end{pmatrix}
\end{align}
where $P_L$ ($P_R$) denotes the projector on the left (right) QD, the valley splittings $\Delta_L$ and $\Delta_R$ are complex numbers and the matrix is written in the valley basis $(\ket{z}, \ket{\bar{z}})$ consisting of two orthogonal valley states. Without loss of generality, we choose the quantization axis in the valley space such that the valley splitting in the left dot is purely real, i.e. $\Im(\Delta_L)=0$. The valley state influences the g-factor in silicon so that stochastic excitations lead to an increase in decoherence. We do not include a valley-dependence of the g-factor, but we do consider valley-excitations as leakage. The resulting energy spectrum is plotted in Fig.~\ref{fig:valley} (a, b). One can see that the valley-orbit interaction can create new (avoided) crossings depending strongly on the valley phase difference $\Delta \phi = \arg(\Delta_L^\ast \Delta_R)$ (here $\Delta \phi = \arg(\Delta_R)$) between the two QDs. These new crossings pose additional adiabaticity restrictions on the pulse to avoid valley excitations.

For the simulation of the valley splitting model, we perform a pulse optimization and noise simulations similar to the preceding section. In the gradient based optimization, the optimized pulses from the model without valley states serve as initial values in the optimization of the valley model.
We use the fidelity measure from Eq.~\eqref{eq:infid}, because it only relies on the truncated propagator on the computational space. In Fig.~\ref{fig:valley} (c, d), we investigate the performance of a parameter combination that yielded excellent fidelities in the absence of valley splittings. The dependence on the valley splitting magnitude $\vert \Delta_R \vert$ and phase $\phi_R$ in the right QD shows that the fidelity drops to unacceptable values for most of the valley splitting configurations. Our strong-driving method only works for a very small phase difference, when the pulse is also adiabatic with respect to the valley DOF. This requirement is unlikely to be fulfilled reliably in practice. A simulation without noise can be found in Fig.~\ref{fig:CrootBenitoCos}~(c, d).

In the strong-driving mode, we are certain to pass all (avoided) crossings in the energy spectrum, but we want to discuss the prospect of avoiding crossings in the weak driving mode by appropriate choice of the working point. A simulation of the valley model in the weak driving regime with the cosine pulse shape as shown in Fig.~\ref{fig:pulses}(a) is plotted in Fig.~\ref{fig:weak_drive_valley}. The simulated parameter set was extracted from an experimental realization of the flopping-mode qubit \cite{crootflopping}. The result shows a strong dependence on the valley splitting configuration with local extrema. The simulation results agree with the experimental results in \cite{crootflopping} if we assume a favorable valley splitting configuration.

The features in the infidelity in Fig.~\ref{fig:weak_drive_valley} can be explained with the corresponding energy spectrum of the valley model. The infidelity contributions at zero valley phase difference $\phi_R = 0$ and $\vert \Delta_R \vert \approx \SI{18}{\micro \electronvolt}$ or  $\vert \Delta_R \vert \approx \SI{80}{\micro \electronvolt}$ are linked to harmonic excitations, where the excitation energy of the driven transition equals the splitting of other transitions (see Appendix~\ref{app::harmonicexcitation} for details). The blue dash-dotted line marks valley configurations where the first two excited states are degenerate in energy, and we need to redefine the computational space above this line to drive spin-like transitions. Each feature that increases the infidelity does not only depend on the sample-depending valley splitting but also on variable parameters like the pulse parameters or magnetic field values. The presence of the valley DOF thus necessitates an individual optimization of the pulse. 

In conclusion, the presence of the valley DOF complicates the driving of the flopping-mode qubit. The expected performance relies on a favorable valley splitting configuration, and recent studies indicate that the valley splitting in the two dots of a DQD are random and uncorrelated \cite{Gefluctuations}. When flopping-mode qubits are fabricated in large numbers in a silicon-based quantum processor, we predict that a relevant fraction will not outperform conventional EDSR driving due to an unfavorable valley splitting configuration. This fraction could be reduced if the processor can vary the physical positions of the DQDs used for flopping-mode qubits and is calibrated for the valley splitting. Also a fraction of unusable flopping-mode qubit locations could be mitigated by the implementation of redundant flopping-mode qubit locations. Another prospect is the research in fabrication and tuning methods to control the valley splitting \cite{TunableValleySplitting, GeSpikeValley, Gefluctuations, GeOscillations}.

\section{Summary and Outlook}
\label{sec::sumandoutlook}

We investigated the performance of the flopping-mode qubit with quantum dynamic simulations and pulse optimizations using the optimal control package qopt \cite{qoptPaper} and publish the source code on the qopt-applications repository \cite{qopt-applications} simultaneously to the article. We demonstrated that driving the flopping-mode qubit with a large pulse amplitude decreases the noise susceptibility and achieves high Rabi frequencies at low magnetic field gradients with the strong displacement of an electron in a DQD. The simulations indicate excellent fidelities even in presence of strong noise sources. In direct comparison to conventional EDSR, the fidelity can be improved by more than two orders of magnitude.

The noise robustness can be attributed to the lower magnetic field gradients and the discreteness of the possible dot positions in between inter-dot transitions in the DQD and this discreteness can be fully leveraged by an appropriate choice of the pulse. We developed a smoothed rectangular pulse form to optimize the manipulation in the strong driving regime. To evaluate the suitability of this pulse form and guide the choice of parameters, we discussed the features of the pulse discreteness and the duty cycle tuning on the qubit performance. Our simulations indicate that the rectangular pulse can further increase the noise insensitivity and the Rabi frequency. In addition, the rectangular pulse can reduce the power dissipation and requirements on the pulse generation.

Leakage caused by orbital excitation due to LZ transitions proved to be one limitation of the rectangular pulse scheme in the strong-driving regime. To avoid diabatic transitions, a small Zeemann splitting $E_Z$ and a strong tunnel coupling $\tunnelcoupling$ are required, while the pulse steepness $\tanhRange$ shows a tradeoff between more adiabatic transitions and an increased noise sensitivity. The flopping-mode qubit makes effective use of the small magnetic field gradient and reaches Rabi frequencies of more than $\frabi/2\pi = \SI{60}{\mega \hertz}$ for realistic magnetic field gradients.

Extending the model by a valley DOF, which is present in silicon based devices, can seriously deteriorate the qubit performance in case of an unfavorable valley splitting configuration. This valley splitting is a material parameter and difficult to predict or manipulate after the fabrication. A pulse optimization including amplitude and offset depending on a given valley splitting is required. Furthermore, the possible drive amplitude is stongly limited.

A further improvement of the simulation could be the inclusion of orbital and valley relaxation as $T_1$-process, by solving a Lindblad master equation of the form
\begin{align}
    \dot{\rho} = - \frac{i}{\hbar} [H_V, \rho] + \sum_{i \in \{O, V \}} \gamma_i \left( L_i \rho L_i^\dagger - \frac{1}{2} \left\{L_i^\dagger L_i, \rho \right\} \right),
\end{align}
where $\rho$ is the systems density matrix, $H_V$ the valley Hamiltonian defined in Eq.~\eqref{eq:HamiltonianValley}, $\gamma_i$ are the relaxation rates and $L_i$ the relaxation operators on the orbital or valley space. The orbital relaxation rate depends mainly on the detuning $\gamma_O(\detuning(t))$ and the valley relaxation rate additionally on the valley splittings $\gamma_V(\detuning, \Delta_L, \Delta_R)$. The correct treatment of the relaxation processes could reduce the population of leakage states and thus increase the fidelity and make the simulation more accurate in settings where leakage cannot be avoided. 
Another noise source to be included is the stability of the micro-magnets at the low total magnetic fields \cite{NeumannMicroMagnet}. 

As the research on the valley splitting progresses and new insights on the distribution of valley splittings are obtained, it would become interesting to simulate the expected infidelity distribution for a given valley splitting distribution. This distribution can then be optimized with an included valley-dependent optimization of the pulse amplitude, offset and envelope or the application of more advanced quantum optimal control methods for example to avoid leakage \cite{ShortcutsToAdiabaticity}.

\begin{acknowledgments}
We thank Lotte Geck for helpful discussions on the feasibility of creating smoothed rectangular pulses with cryoelectronics. We thank Lars Schreiber and Max Oberländer for proof reading the manuscript. This work has been funded by the Federal Ministry of Education and Research (Germany), Funding code 13N15652 (QUASAR). We acknowledge support from the Impulse and Networking Fund of the Helmholtz Association.

\end{acknowledgments}

\textbf{Author's contribution:} F. Butt performed preliminary studies under the supervision of J. Teske, P. Cerfontaine and H. Bluhm. J. Teske designed the pulse schemes, developed the optimization algorithm, programmed the simulations in this manuscript, analyzed the data, and wrote the manuscript with guidance by G. Burkard and H. Bluhm.

\textbf{Conflicts of interest:} H. Bluhm is co-inventor of patent applications that cover the flopping mode EDSR as driving mode.

\appendix
\section{Pulse optimization}
\label{app::optimizaiton}

In this section, we present the construction details of the rectangular pulse scheme using analytic and numeric pulse optimization. We start with a discussion of the degrees of freedom of the control pulses and the target operation. Then we describe the time domain optimization with the qopt package and how the time discretization is chosen. Finally we explain the actual optimization algorithm for the cosine pulse and the more complicated case of the rectangular pulse, which requires an accurate estimation of the achievable Rabi frequency and the effective resonance frequency for the optimization of the discrete pulse time.

There are two relevant degrees of freedom for a quantum gate created by Rabi driving. Using the isometry of SU(2) and SO(3), any target evolution of our qubit can be described as rotation on the Bloch sphere with a rotation angle $\phi$ and a rotation axis. We require the rotation axis to lay in the xy-plane so that the axis can be described by its azimuth angle $\theta$ in polar coordinates.

The azimuth angle $\theta$ can be tuned using a phase shift of the control pulses. This tuning step is the simplest, because any phase shift in the control pulse directly corresponds to a shift in $\theta$. To reduce the number of optimization parameters and the cost functions of the optimization, we start with an optimization of the rotation angle $\phi$. This is effectively achieved by choosing a state infidelity as the cost function. Thereby, we assume our qubit to be initially in the ground state and drive a full $\pi$-rotation to the first excited state. Since we assume the orbital splitting to be larger than the Zeemann splitting, this is a spin-like transition.

We perform the simulation and optimization with the optimal control package qopt making use of the support of analytic gradients. For this purpose we only need to implement our pulse parameterization and derivatives of this paramterization.

Our pulses possess four degrees of freedom being the amplitude, frequency, length and phase shift, but in the strong driving regime not all parameters can be used effectively, because the dynamics are insensitive to changes of the amplitude. Further, the pulse frequency must be chosen resonant to the spin precession frequency and this choice must be consistent with the pulse, because the effective average Zeeman splitting depends on the electron position $\Delta B_z\neq 0$. The phase shift is used to control the azimuth angle $\theta$ of the quantum gate. The pulse length as remaining DOF can be optimized to control the rotation angle $\phi$.

The optimal control package qopt operates with fixed time steps but we can use a scalar time stretching parameter $\timestrech$ to introduce an effective optimization of the pulse length. This time stretching parameter is absorbed into the Hamiltonian using the identity
\begin{align}
    U = e^{-i \hbar (t \timestrech) H} = e^{-i \hbar t (\timestrech H)}.
\end{align}

\subsubsection{Time Discretization}
A crucial step in the optimization is the choice of the time discretization into time steps of length $\delta t$, because a too long $\delta t$ can limit the accuracy, while a too short $\delta t$ decreases the numerical efficiency. We thus need to identify the longest $\delta t$ that resolves all relevant dynamics of our system.
Only the orbital dynamics are relevant for the choice of $\delta t$, because they are much faster than the spin dynamics. Let's consider the orbital Hamiltonian
\begin{align}
    H_O(\detuning(t)) = \frac{\detuning(t)}{2}\sigma_z + \tunnelcoupling \sigma_x
\end{align}
and two unitary propagators $U_1, U_2$ with different time discretizations
\begin{align*}
    U_1 &= \prod_{k=1}^2 e^{H(\epsilon(k\delta t))\delta t}\\
    U_2 &= e^{H(\epsilon(0)) 2 \delta t},
\end{align*}
such that $U_1$ is sampled in two time steps of $\delta t$ and $U_2$ is sampled with one time step $2\delta t$. Next, we estimate how small $\delta t$ needs to be chosen that the discretization in steps of $\delta t$ and $2 \delta t$ yield approximately the same result.
Therefore, we calculate the deviation $\Delta U$ of these two propagators with the Baker-Campbell-Hausdorff formula to be
\begin{align}
    \Delta U = U_1 U_2^\dagger \approx \exp\{(H(\epsilon(0))-H(\epsilon(\delta t))) \delta t \label{eq:condition1timedisc} \\
    - 2 \delta t^2 [H(\epsilon(0)), H(\epsilon(\delta t))]\}. \label{eq:condition2timedisc} 
\end{align}
We can now identify criteria that ensure $\Delta U$ to be close to unity.
From \eqref{eq:condition1timedisc} we identify the first condition to be 
\begin{align}
    \delta \detuning \delta t \ll 1
\end{align}
with $\delta \detuning = (\detuning(0) - \detuning(\delta t))/2$, and calculating the commutator in \eqref{eq:condition2timedisc} gives the second condition:
\begin{align}
    4\tunnelcoupling \delta t^2 \delta \detuning \ll 1.
\end{align}
Using the maximum values of the pulse derivatives by time as bound for $\delta \detuning \approx \delta t \frac{\partial \detuning}{\partial t}$, we can reformulate the conditions for both pulses in terms of the number of time steps $n_t$ and the total pulse time $T$:

\begin{center}
\begin{tabular}{ c|c|c }
& Cosine & Rectangular \\ \hline
C 1 & $n_t \gg \sqrt{T^2 E_z A}$ & $n_t \gg \sqrt{T^2 A \tanhRange \frac{E_z}{\pi}}$\\ \hline
C 2 & $n_t \gg (T^3 4 \tunnelcoupling E_z A)^{1/3}$ & $n_t \gg (T^3 4 \tunnelcoupling A \tanhRange \frac{E_z}{\pi})^{1/3}$
\end{tabular}
\end{center}

\subsection{Cosine Pulse Optimization}

For the cosine pulse, we optimize the time stretching parameter $\timestrech$ and the pulse frequency $\omega$. Using $\timestrech$ to optimize the length of the cosine pulse, we need to compensate changes in $\omega$ to stay in resonance with the Zeeman splitting. The amplitude and phase remain constant during the optimization. 

We calculate initial values for the optimization from the Rabi driving theory \cite{RabiOriginal}. First, we transform the time-dependend Hamiltonian $H(\detuning(t))$ into the eigenbasis of the Hamiltonian at the central position $H(\detuning_0)$ with the basis transformation $V(\detuning_0)$. The eigenvalues of 
\begin{align}
    V(\detuning_0)H(\detuning_0)V(\detuning_0)^\dagger = \text{diag}(E_1, E_2, E_3, E_4)
\end{align}
are sorted in ascending order $E_1 \leq E_2 \leq E_3 \leq E_4$. We want to drive the spin-like transition of resonance frequency $\omres = E_2 - E_1$, while the electron remains in the orbital ground state. The Rabi frequency $\frabi$ can be approximated by the off-diagonal element corresponding to the transition \cite{RabiBenjamin}:
\begin{align}
    \frabi \approx A \left(\frac{\partial H(\detuning)}{\partial \detuning}\vert_{\detuning_0}\right)_{[0, 1]} \approx (H(A) - H(-A))_{[0, 1]}.
\end{align}

\subsection{Rectangular Pulse Optimization}

Tuning the smoothed rectangular pulse is more complicated, since we require the pulse to consist of a number $N_P$ of full driving periods. In addition, we want to avoid a mixed discrete and continuous optimization for the sake of simplicity, so we require $N_P$ to be constant during the continuous optimization. In consequence, the pulse length and frequency cannot be varied independently. We calculate the optimal number of driving periods before the optimization and optimize only the time stretching parameter $\timestrech$ and the duty cycle parameter $\cdc$.

The duty cycle tuning can be used to effectively decrease the driving strength and thereby change the Rabi frequency $\frabi$. $\frabi$ is maximal when $\cdc=0$ and drops to zero for $\cdc=1$. We want to choose $\cdc$ as small as possible, because large $\cdc$ values leads to longer pulses and an increased noise susceptibility. Therefore the optimal $N_p$ is the smallest number of driving periods that are sufficient to realise the target gate.

\subsubsection{Rabi frequency Estimation}
We calculate an estimation of the maximal $\frabi$ at $\cdc=0$ by solving the effective spin Hamiltonian
\begin{align}
    H_S = E_z \frac{\sigma_z}{2} + \Delta B_x \langle \tau_z \rangle  \frac{\sigma_x}{2}
\end{align}
where $< \tau_z> = \langle 0 \vert \tau_z \vert 0 \rangle$ is the expectation value of the $\tau_z$ operator in the instantaneous ground state. This expectation value can be calculated from the orbital Hamiltonian:
\begin{align}
    H_O = \detuning \frac{\tau_z}{2} + 2 \tunnelcoupling \frac{\tau_x}{2}
\end{align}
and has the analytical solution
\begin{align}
    \langle \tau_z \rangle = \frac{\detuning (\detuning - \sqrt{\detuning^2 + 4 \tunnelcoupling^2})}{\detuning^2 - \detuning \sqrt{\detuning^2 + 4 \tunnelcoupling^2 } + 4 \tunnelcoupling^2 }.
\end{align}
Finally, we calculate the Rabi frequency $\frabi$ in the rotating reference frame as the average over one period of the driving pulse:
\begin{align}
    \omega_R = \Delta B_x \frac{\sigma_x}{\Tres} \int^{\Tres}_0 \langle \tau_z(\detuning(t)) \rangle dt.
\end{align}
The discreteness of the pulse leads to systematic errors even after the optimization. These errors lead only to small infidelities but are visible for example as a vertical line in Fig.~\ref{fig:LZSandFastPulse} (a, b) around $\tanhRange=8$. At this line, $N_P$ changed by one.

Furthermore, precise initial values are required for the optimization, because the LZS-oscillations (see appendix \ref{app::LZS}) create such a rough optimization landscape that a pure gradient based optimization quickly converges to the nearest local optimum. We therefore start by simulating a grid of parameter combinations around the presumed location of the global optimum. The parameter combination with the lowest infidelity is then chosen as the initial set of values.

\subsubsection{Effective Zeeman Splitting}

We further improve the convergence of the optimization by introduction of orthogonal control parameters. $\cdc$ and $\timestrech$ are not orthogonal, in the sense that any change in $\cdc$ alters the average occupation of the QDs and changes the effective resonance frequency $\omres$ because 
of the longitudinal magnetic field gradient $\Delta B_z$ in the Hamiltonian in \eqref{eq:hamiltonian}. We now calculate the shift in $\omres$ and the corresponding compensation in $\timestrech$ analytically and include it in the pulse parameterization.

For any duty cycle parameter $\cdc$, we spend $\cdc / 2  \cdot \Tres$ longer in the right QD. For this time, the resonance frequency is shifted by $\Delta B_z$. Thus the new effective resonance frequency is $\omres^\prime = E_z(1 + \frac{\cdc \Delta B_z}{2 E_z})$, and we need to multiply $\timestrech$ by an additional correction of $s_C = 1 / (1 + \frac{\cdc \Delta B_z}{2 E_z})\approx 1 - \frac{\cdc \Delta B_z}{2 E_z}$.

\section{Landau-Zener-Stückelberg Interference}
\label{app::LZS}

\begin{figure}
    \centering
    \includegraphics{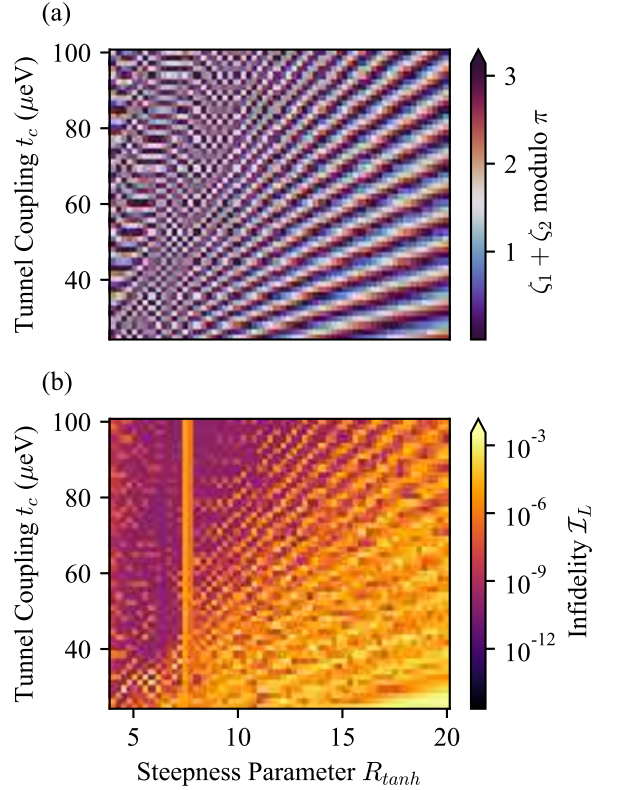}
    \caption{Landau-Zener-Stückelberg oscillation pattern. (a) Shows the sum of the accumulated phases $\zeta_1 + \zeta_2$ between transition through the avoided crossing modulo $2\pi$ for the simulation in Fig.~\ref{fig:LZSandFastPulse} (a,b) in the main text. The pattern can also be found in the infidelity calculated in the absence of noise plotted in (b).}
    \label{fig:LZSPhaseOscillation}
\end{figure}

In Fig.~\ref{fig:LZSandFastPulse} (a, b) in the main text, we can weakly see an oscillation pattern in the infidelity. This pattern can be explained by Landau-Zener-Stückelberg (LZS) interferometry \cite{LZS_SHEVCHENKO20101}. 
The LZS theory describes coherent excitations that occur as result of periodic passages of an avoided crossing. Driving the detuning induces such passages in the orbital DOF in the flopping-mode qubit, where the avoided crossing is at the minimum of the orbital splitting at $\detuning=0$ and excitations correspond to leakage into the excited orbital state. 
The resonance condition for LZS interferometry is linked to the phases $\zeta_1$ and $\zeta_2$ accumulated between the passages of the avoided crossing. In the slow-passage limit, the resonance condition can be reduced to 
\begin{align}
    \zeta_1 + \zeta_2 + 2\phi_S = k\pi, k \in \mathbb{N},
\end{align}
where $\phi_S$ is the Stückelberg phase. The correspondence between the resonance pattern in the infidelity in Fig.~\ref{fig:LZSandFastPulse}~(a) or Fig.~\ref{fig:LZSPhaseOscillation}~(b) and the sum of the accumulated phases in Fig.~\ref{fig:LZSPhaseOscillation}~(a) indicates that LZS-oscillations pose the main contribution of infidelity in the absence of noise in this simulation.

\section{Harmonic Excitation}
\label{app::harmonicexcitation}

\begin{figure}
    \centering
    \includegraphics{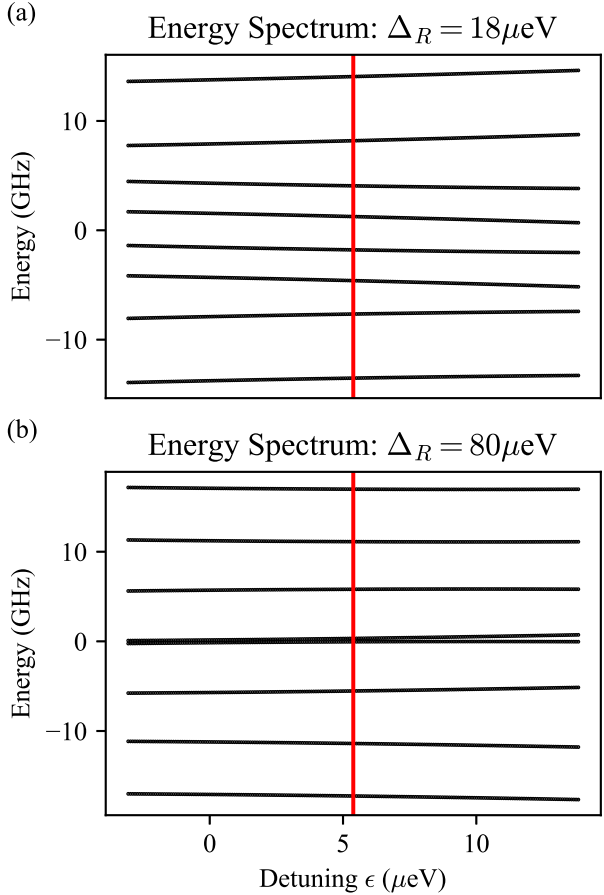}
    \caption{Selected energy spectra with parameters of the weak-driving mode with valley DOF as plotted in Fig.~\ref{fig:weak_drive_valley}. In both plots we discuss energy spectra for purely real valley splittings in both dots, i.e., $\Delta \phi = 0$. The red lines mark the operation point in the pulse offset $\detuning_0$ and the plotted range corresponds to $10 A$. The energy spectra can explain leakage into certain states. In (a) the energy difference between the ground and first excited state equals approximately the energy splitting between the first and the third excited state. In (b) the energy difference between the ground the the first excited state equal approximately the energy difference between the first and the second excited state.}
    \label{fig:weak_harmonic}
\end{figure}

This section discusses the energy spectrum of valley splitting model to explain features in Fig.~\ref{fig:weak_drive_valley}. For the plots in Fig.~\ref{fig:weak_harmonic}, we select two points of interest from the simulation with a purely real valley parameter $\Delta_R = \SI{18}{\micro \electronvolt}$ in (a) and $\Delta_R = \SI{80}{\micro \electronvolt}$ in (b). At these points we observe a local increase in the infidelity in Fig.~\ref{fig:weak_drive_valley} that cannot be explained with diabatic valley excitation, because the valley phase difference is zero.

We explain these features in the infidelity by harmonic excitation, in the sense that multiple transitions have an energy splitting identical to the driven transition. The Hamiltonian $H_V(\detuning_0)$ from Eq.~\eqref{eq:HamiltonianValley} has the eigenvalues $e_1, \dots, e_8$ in ascending order and eigenvectors $v_1, \dots, v_8$, where $v_1$ and $v_2$ span the computational space. Thus our pulse is driven with the frequency $e_2 - e_1 = \omega \approx E_z$. Now we observe an increase in the Leakage into state $v_4$ for $\Delta_R=\SI{18}{\micro \electronvolt}$ in our simulation. The reason for this increase in leakage is that $e_2 - e_1 = \omega \approx e_4 - e_2$ as can be seen in Fig.~\ref{fig:weak_harmonic}~(a). Similarly, we observe leakage into $v_3$ around $\Delta_R=\SI{18}{\micro \electronvolt}$ and Fig.~\ref{fig:weak_harmonic}~(b) shows that here $\omega \approx e_3 - e_2$ holds. We conclude that the observed leakage increases at these points, because our pulse drives multiple transitions resonantly with the same energy difference.

\section{Reproduction Croot et al.}
\label{app::croot}

\begin{figure*}
    \centering
    \includegraphics{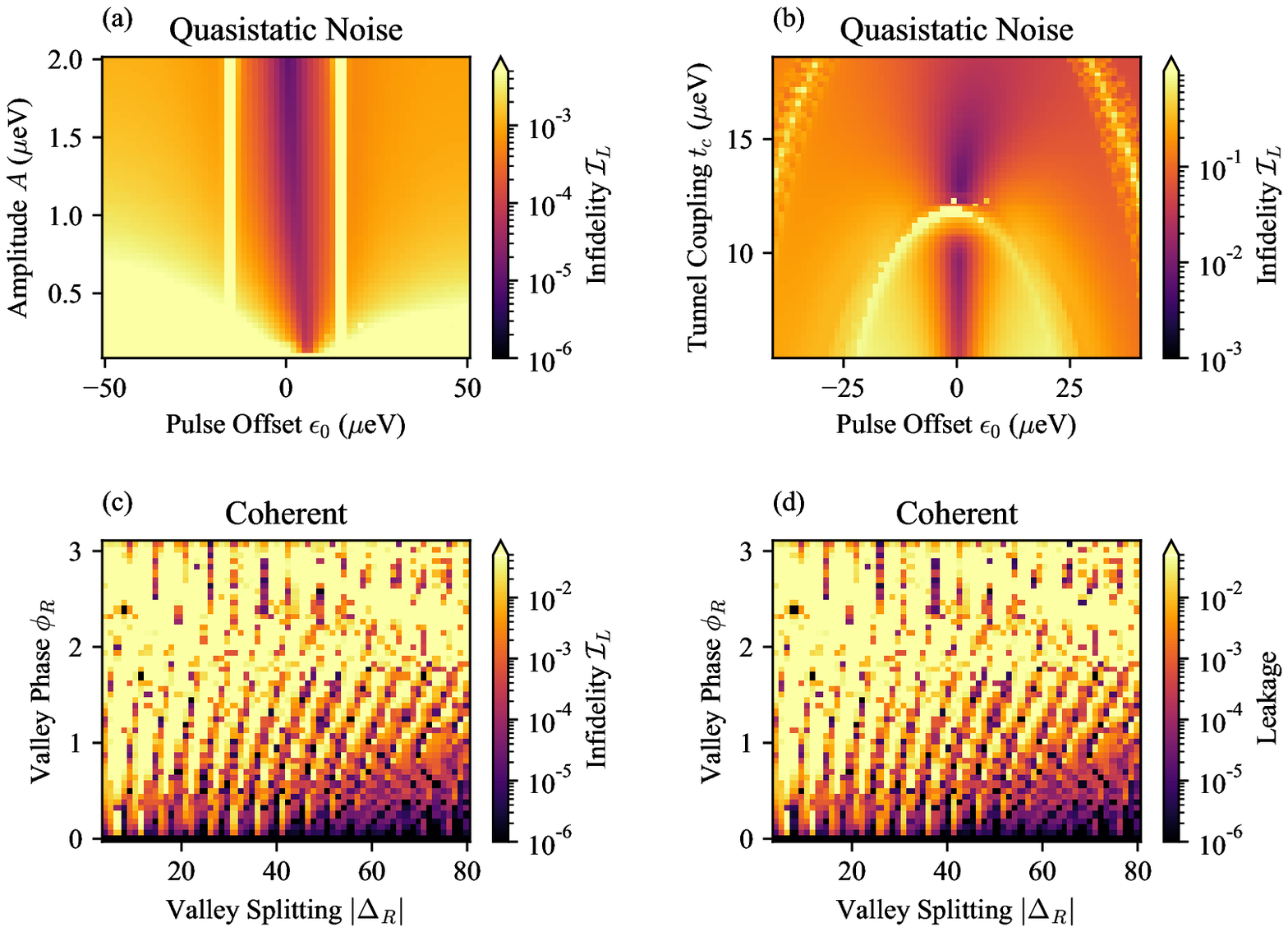}
    \caption{(a) Reproduction of the infidelity in Fig.~4~(b) in \cite{crootflopping} measured by Croot \textit{et al.}. 
    The simulation was performed with the parameters measured in \cite{crootflopping}. We therefore used magnetic field gradients of $\Delta B_x = \SI{15}{\milli \tesla}$ and $\Delta B_z= \SI{0.27}{\milli \tesla}$, a total magnetic field of $ B_z = \SI{209.4}{\milli \tesla}$, a tunnel coupling of $2\tunnelcoupling =\SI{23}{\micro \electronvolt}$ and quasistatic noise with a standard deviation of $\sigma_\detuning =\SI{0.5}{\micro \electronvolt}$. 
    (b) Reproduction of the infidelity in Fig.~4~(b) in \cite{benitoflopping} simulated by Benito \textit{et al.}. A second order charge noise sweet spot at about $\tunnelcoupling=\SI{13}{\micro \electronvolt}$ can be seen, but the minimum is shifted towards the center $\detuning=0$. The simulation was performed with the parameters employed in \cite{crootflopping}. We therefore used magnetic field gradients $\Delta B_x = \SI{17.3}{\milli \tesla}$ and $\Delta B_z  = \SI{4.32}{\milli \tesla}$ at a total magnetic field of $ B_z= \SI{207}{\milli \tesla}$, driving with an amplitude of $A=\SI{2.1}{\micro \electronvolt}$ and simulating quasistatic noise of standard deviation $\sigma_\detuning /\hbar=\SI{2.5}{\micro \electronvolt}$.
    (c, d) show the coherent simulation of the flopping-mode qubit including the valley splitting but without noise. The simulation was performed with the same parameters as in Fig.~\ref{fig:valley}. The Infidelity (c) and the Leakage (d) look almost identical, indicating that leakage is the dominant contribution to the infidelity.
    }
    \label{fig:CrootBenitoCos}
\end{figure*}

Croot \textit{et al.} demonstrated the feasibility of flopping-mode driving experimentally \cite{crootflopping}. They leveraged the large electric dipole moment to increase the driving efficiency by nearly three orders of magnitude in the zero detuning configuration. This enabled them to reach Rabi-frequencies of up to $\SI{8}{\mega \hertz}$ with about 250 times less microwave power compared to standard EDSR.

We qualitatively reproduce Fig.~4~(b) from \cite{crootflopping} by sweeping the static offset $\detuning_0$ and the driving amplitude $A$. Our results are plotted in Fig.~\ref{fig:CrootBenitoCos}~(a) and the simulation reproduces the fidelity maximum near zero detuning offset $\detuning_0=0$ with a slight shift to positive detuning offsets. Croot \textit{et al.} do not find the local infidelity maximum at the spin-orbit degeneracy point, possibly because they sample the offset $\detuning_0$ too coarsely or because other dephasing channels dominate. We cannot reproduce their high Rabi-frequencies for large detuning offsets $\vert\detuning\vert>\SI{30}{\micro \electronvolt} $, likely because the electron is strongly confined for these values such that the driving relies on shifts of the QD, which are not included in our model.

\section{Reproduction Benito et al.}
\label{app::benito}

Benito \textit{et al.} investigated the susceptibility of the flopping-mode qubit to electric noise with analytical calculations. They predicted a second order sweet spot for the qubit, where it would be insensitive to quasi static detuning shifts up to second order. We can reproduce a local minimum in the infidelity but it is shifted more towards the symmetric position $\detuning=0$. The differences between Fig.~\ref{fig:CrootBenitoCos}~(b) and Fig.~4~(b) in \cite{benitoflopping} can be explained by the fact that the numerical simulations capture the influence of leakage explicitly than the analytical calculations employed in \cite{benitoflopping}.

The main difference between the driving mode investigated in \cite{crootflopping, benitoflopping} and the driving mode presented in this manuscript are the relations of magnetic field strengths, tunnel coupling and driving amplitude. They operate at a much larger spin-orbit mixing and achieve a better relation of Rabi frequency to driving power at the cost of an increased noise susceptibility.

\bibliography{bibliography}

\begin{thebibliography}{49}%
\makeatletter
\providecommand \@ifxundefined [1]{%
 \@ifx{#1\undefined}
}%
\providecommand \@ifnum [1]{%
 \ifnum #1\expandafter \@firstoftwo
 \else \expandafter \@secondoftwo
 \fi
}%
\providecommand \@ifx [1]{%
 \ifx #1\expandafter \@firstoftwo
 \else \expandafter \@secondoftwo
 \fi
}%
\providecommand \natexlab [1]{#1}%
\providecommand \enquote  [1]{``#1''}%
\providecommand \bibnamefont  [1]{#1}%
\providecommand \bibfnamefont [1]{#1}%
\providecommand \citenamefont [1]{#1}%
\providecommand \href@noop [0]{\@secondoftwo}%
\providecommand \href [0]{\begingroup \@sanitize@url \@href}%
\providecommand \@href[1]{\@@startlink{#1}\@@href}%
\providecommand \@@href[1]{\endgroup#1\@@endlink}%
\providecommand \@sanitize@url [0]{\catcode `\\12\catcode `\$12\catcode
  `\&12\catcode `\#12\catcode `\^12\catcode `\_12\catcode `\%12\relax}%
\providecommand \@@startlink[1]{}%
\providecommand \@@endlink[0]{}%
\providecommand \url  [0]{\begingroup\@sanitize@url \@url }%
\providecommand \@url [1]{\endgroup\@href {#1}{\urlprefix }}%
\providecommand \urlprefix  [0]{URL }%
\providecommand \Eprint [0]{\href }%
\providecommand \doibase [0]{https://doi.org/}%
\providecommand \selectlanguage [0]{\@gobble}%
\providecommand \bibinfo  [0]{\@secondoftwo}%
\providecommand \bibfield  [0]{\@secondoftwo}%
\providecommand \translation [1]{[#1]}%
\providecommand \BibitemOpen [0]{}%
\providecommand \bibitemStop [0]{}%
\providecommand \bibitemNoStop [0]{.\EOS\space}%
\providecommand \EOS [0]{\spacefactor3000\relax}%
\providecommand \BibitemShut  [1]{\csname bibitem#1\endcsname}%
\let\auto@bib@innerbib\@empty
\bibitem [{\citenamefont {Noiri}\ \emph
  {et~al.}(2022{\natexlab{a}})\citenamefont {Noiri}, \citenamefont {Takeda},
  \citenamefont {Nakajima}, \citenamefont {Kobayashi}, \citenamefont {Sammak},
  \citenamefont {Scappucci},\ and\ \citenamefont
  {Tarucha}}]{Noiri2022HighFidGates}%
  \BibitemOpen
  \bibfield  {author} {\bibinfo {author} {\bibfnamefont {A.}~\bibnamefont
  {Noiri}}, \bibinfo {author} {\bibfnamefont {K.}~\bibnamefont {Takeda}},
  \bibinfo {author} {\bibfnamefont {T.}~\bibnamefont {Nakajima}}, \bibinfo
  {author} {\bibfnamefont {T.}~\bibnamefont {Kobayashi}}, \bibinfo {author}
  {\bibfnamefont {A.}~\bibnamefont {Sammak}}, \bibinfo {author} {\bibfnamefont
  {G.}~\bibnamefont {Scappucci}},\ and\ \bibinfo {author} {\bibfnamefont
  {S.}~\bibnamefont {Tarucha}},\ }\bibfield  {title} {\bibinfo {title} {Fast
  universal quantum gate above the fault-tolerance threshold in silicon},\
  }\href {https://doi.org/10.1038/s41586-021-04182-y} {\bibfield  {journal}
  {\bibinfo  {journal} {Nature}\ }\textbf {\bibinfo {volume} {601}},\ \bibinfo
  {pages} {338} (\bibinfo {year} {2022}{\natexlab{a}})}\BibitemShut {NoStop}%
\bibitem [{\citenamefont {Xue}\ \emph {et~al.}(2022)\citenamefont {Xue},
  \citenamefont {Russ}, \citenamefont {Samkharadze}, \citenamefont {Undseth},
  \citenamefont {Sammak}, \citenamefont {Scappucci},\ and\ \citenamefont
  {Vandersypen}}]{Xue2022HighFidGates}%
  \BibitemOpen
  \bibfield  {author} {\bibinfo {author} {\bibfnamefont {X.}~\bibnamefont
  {Xue}}, \bibinfo {author} {\bibfnamefont {M.}~\bibnamefont {Russ}}, \bibinfo
  {author} {\bibfnamefont {N.}~\bibnamefont {Samkharadze}}, \bibinfo {author}
  {\bibfnamefont {B.}~\bibnamefont {Undseth}}, \bibinfo {author} {\bibfnamefont
  {A.}~\bibnamefont {Sammak}}, \bibinfo {author} {\bibfnamefont
  {G.}~\bibnamefont {Scappucci}},\ and\ \bibinfo {author} {\bibfnamefont
  {L.~M.~K.}\ \bibnamefont {Vandersypen}},\ }\bibfield  {title} {\bibinfo
  {title} {Quantum logic with spin qubits crossing the surface code
  threshold},\ }\href {https://doi.org/10.1038/s41586-021-04273-w} {\bibfield
  {journal} {\bibinfo  {journal} {Nature}\ }\textbf {\bibinfo {volume} {601}},\
  \bibinfo {pages} {343} (\bibinfo {year} {2022})}\BibitemShut {NoStop}%
\bibitem [{\citenamefont {Philips}\ \emph {et~al.}(2022)\citenamefont
  {Philips}, \citenamefont {Madzik}, \citenamefont {Amitonov}, \citenamefont
  {de~Snoo}, \citenamefont {Russ}, \citenamefont {Kalhor}, \citenamefont
  {Volk}, \citenamefont {Lawrie}, \citenamefont {Brousse}, \citenamefont
  {Tryputen}, \citenamefont {Wuetz}, \citenamefont {Sammak}, \citenamefont
  {Veldhorst}, \citenamefont {Scappucci},\ and\ \citenamefont
  {Vandersypen}}]{6qubitPhilips2022}%
  \BibitemOpen
  \bibfield  {author} {\bibinfo {author} {\bibfnamefont {S.~G.~J.}\
  \bibnamefont {Philips}}, \bibinfo {author} {\bibfnamefont {M.~T.}\
  \bibnamefont {Madzik}}, \bibinfo {author} {\bibfnamefont {S.~V.}\
  \bibnamefont {Amitonov}}, \bibinfo {author} {\bibfnamefont {S.~L.}\
  \bibnamefont {de~Snoo}}, \bibinfo {author} {\bibfnamefont {M.}~\bibnamefont
  {Russ}}, \bibinfo {author} {\bibfnamefont {N.}~\bibnamefont {Kalhor}},
  \bibinfo {author} {\bibfnamefont {C.}~\bibnamefont {Volk}}, \bibinfo {author}
  {\bibfnamefont {W.~I.~L.}\ \bibnamefont {Lawrie}}, \bibinfo {author}
  {\bibfnamefont {D.}~\bibnamefont {Brousse}}, \bibinfo {author} {\bibfnamefont
  {L.}~\bibnamefont {Tryputen}}, \bibinfo {author} {\bibfnamefont {B.~P.}\
  \bibnamefont {Wuetz}}, \bibinfo {author} {\bibfnamefont {A.}~\bibnamefont
  {Sammak}}, \bibinfo {author} {\bibfnamefont {M.}~\bibnamefont {Veldhorst}},
  \bibinfo {author} {\bibfnamefont {G.}~\bibnamefont {Scappucci}},\ and\
  \bibinfo {author} {\bibfnamefont {L.~M.~K.}\ \bibnamefont {Vandersypen}},\
  }\href {https://doi.org/10.48550/ARXIV.2202.09252} {\bibinfo {title}
  {Universal control of a six-qubit quantum processor in silicon}} (\bibinfo
  {year} {2022}),\ \Eprint {https://arxiv.org/abs/2202.09252}
  {arXiv:2202.09252} \BibitemShut {NoStop}%
\bibitem [{\citenamefont {Mills}\ \emph {et~al.}(2022)\citenamefont {Mills},
  \citenamefont {Guinn}, \citenamefont {Gullans}, \citenamefont {Sigillito},
  \citenamefont {Feldman}, \citenamefont {Nielsen},\ and\ \citenamefont
  {Petta}}]{2qubitprocessorHighfid2022}%
  \BibitemOpen
  \bibfield  {author} {\bibinfo {author} {\bibfnamefont {A.~R.}\ \bibnamefont
  {Mills}}, \bibinfo {author} {\bibfnamefont {C.~R.}\ \bibnamefont {Guinn}},
  \bibinfo {author} {\bibfnamefont {M.~J.}\ \bibnamefont {Gullans}}, \bibinfo
  {author} {\bibfnamefont {A.~J.}\ \bibnamefont {Sigillito}}, \bibinfo {author}
  {\bibfnamefont {M.~M.}\ \bibnamefont {Feldman}}, \bibinfo {author}
  {\bibfnamefont {E.}~\bibnamefont {Nielsen}},\ and\ \bibinfo {author}
  {\bibfnamefont {J.~R.}\ \bibnamefont {Petta}},\ }\bibfield  {title} {\bibinfo
  {title} {Two-qubit silicon quantum processor with operation fidelity
  exceeding 99\%},\ }\href {https://doi.org/10.1126/sciadv.abn5130} {\bibfield
  {journal} {\bibinfo  {journal} {Science Advances}\ }\textbf {\bibinfo
  {volume} {8}},\ \bibinfo {pages} {eabn5130} (\bibinfo {year} {2022})},\
  \Eprint
  {https://arxiv.org/abs/https://www.science.org/doi/pdf/10.1126/sciadv.abn5130}
  {https://www.science.org/doi/pdf/10.1126/sciadv.abn5130} \BibitemShut
  {NoStop}%
\bibitem [{\citenamefont {Burkard}\ \emph {et~al.}(2021)\citenamefont
  {Burkard}, \citenamefont {Ladd}, \citenamefont {Nichol}, \citenamefont
  {Pan},\ and\ \citenamefont {Petta}}]{Burkard2021}%
  \BibitemOpen
  \bibfield  {author} {\bibinfo {author} {\bibfnamefont {G.}~\bibnamefont
  {Burkard}}, \bibinfo {author} {\bibfnamefont {T.~D.}\ \bibnamefont {Ladd}},
  \bibinfo {author} {\bibfnamefont {J.~M.}\ \bibnamefont {Nichol}}, \bibinfo
  {author} {\bibfnamefont {A.}~\bibnamefont {Pan}},\ and\ \bibinfo {author}
  {\bibfnamefont {J.~R.}\ \bibnamefont {Petta}},\ }\href@noop {} {\bibinfo
  {title} {Semiconductor spin qubits}} (\bibinfo {year} {2021}),\ \Eprint
  {https://arxiv.org/abs/2112.08863} {arXiv:2112.08863 [cond-mat.mes-hall]}
  \BibitemShut {NoStop}%
\bibitem [{\citenamefont {Langrock}\ \emph {et~al.}(2022)\citenamefont
  {Langrock}, \citenamefont {Krzywda}, \citenamefont {Focke}, \citenamefont
  {Seidler}, \citenamefont {Schreiber},\ and\ \citenamefont
  {Cywiński}}]{BlueprintShuttle}%
  \BibitemOpen
  \bibfield  {author} {\bibinfo {author} {\bibfnamefont {V.}~\bibnamefont
  {Langrock}}, \bibinfo {author} {\bibfnamefont {J.~A.}\ \bibnamefont
  {Krzywda}}, \bibinfo {author} {\bibfnamefont {N.}~\bibnamefont {Focke}},
  \bibinfo {author} {\bibfnamefont {I.}~\bibnamefont {Seidler}}, \bibinfo
  {author} {\bibfnamefont {L.~R.}\ \bibnamefont {Schreiber}},\ and\ \bibinfo
  {author} {\bibfnamefont {L.}~\bibnamefont {Cywiński}},\ }\href
  {https://doi.org/10.48550/ARXIV.2202.11793} {\bibinfo {title} {Blueprint of a
  scalable spin qubit shuttle device for coherent mid-range qubit transfer in
  disordered si/sige/sio$_2$}} (\bibinfo {year} {2022}),\ \Eprint
  {https://arxiv.org/abs/2202.11793} {arXiv:2202.11793} \BibitemShut {NoStop}%
\bibitem [{\citenamefont {Boter}\ \emph {et~al.}(2021)\citenamefont {Boter},
  \citenamefont {Dehollain}, \citenamefont {van Dijk}, \citenamefont {Xu},
  \citenamefont {Hensgens}, \citenamefont {Versluis}, \citenamefont {Naus},
  \citenamefont {Clarke}, \citenamefont {Veldhorst}, \citenamefont
  {Sebastiano},\ and\ \citenamefont {Vandersypen}}]{SpiderWebArray}%
  \BibitemOpen
  \bibfield  {author} {\bibinfo {author} {\bibfnamefont {J.~M.}\ \bibnamefont
  {Boter}}, \bibinfo {author} {\bibfnamefont {J.~P.}\ \bibnamefont
  {Dehollain}}, \bibinfo {author} {\bibfnamefont {J.~P.~G.}\ \bibnamefont {van
  Dijk}}, \bibinfo {author} {\bibfnamefont {Y.}~\bibnamefont {Xu}}, \bibinfo
  {author} {\bibfnamefont {T.}~\bibnamefont {Hensgens}}, \bibinfo {author}
  {\bibfnamefont {R.}~\bibnamefont {Versluis}}, \bibinfo {author}
  {\bibfnamefont {H.~W.~L.}\ \bibnamefont {Naus}}, \bibinfo {author}
  {\bibfnamefont {J.~S.}\ \bibnamefont {Clarke}}, \bibinfo {author}
  {\bibfnamefont {M.}~\bibnamefont {Veldhorst}}, \bibinfo {author}
  {\bibfnamefont {F.}~\bibnamefont {Sebastiano}},\ and\ \bibinfo {author}
  {\bibfnamefont {L.~M.~K.}\ \bibnamefont {Vandersypen}},\ }\href
  {https://doi.org/10.48550/ARXIV.2110.00189} {\bibinfo {title} {The spider-web
  array--a sparse spin qubit array}} (\bibinfo {year} {2021}),\ \Eprint
  {https://arxiv.org/abs/2110.00189} {arXiv:2110.00189} \BibitemShut {NoStop}%
\bibitem [{\citenamefont {Xue}\ \emph {et~al.}(2021)\citenamefont {Xue},
  \citenamefont {Patra}, \citenamefont {van Dijk}, \citenamefont {Samkharadze},
  \citenamefont {Subramanian}, \citenamefont {Corna}, \citenamefont
  {Paquelet~Wuetz}, \citenamefont {Jeon}, \citenamefont {Sheikh}, \citenamefont
  {Juarez-Hernandez}, \citenamefont {Esparza}, \citenamefont {Rampurawala},
  \citenamefont {Carlton}, \citenamefont {Ravikumar}, \citenamefont {Nieva},
  \citenamefont {Kim}, \citenamefont {Lee}, \citenamefont {Sammak},
  \citenamefont {Scappucci}, \citenamefont {Veldhorst}, \citenamefont
  {Sebastiano}, \citenamefont {Babaie}, \citenamefont {Pellerano},
  \citenamefont {Charbon},\ and\ \citenamefont
  {Vandersypen}}]{Xue2021CMOSCryo}%
  \BibitemOpen
  \bibfield  {author} {\bibinfo {author} {\bibfnamefont {X.}~\bibnamefont
  {Xue}}, \bibinfo {author} {\bibfnamefont {B.}~\bibnamefont {Patra}}, \bibinfo
  {author} {\bibfnamefont {J.~P.~G.}\ \bibnamefont {van Dijk}}, \bibinfo
  {author} {\bibfnamefont {N.}~\bibnamefont {Samkharadze}}, \bibinfo {author}
  {\bibfnamefont {S.}~\bibnamefont {Subramanian}}, \bibinfo {author}
  {\bibfnamefont {A.}~\bibnamefont {Corna}}, \bibinfo {author} {\bibfnamefont
  {B.}~\bibnamefont {Paquelet~Wuetz}}, \bibinfo {author} {\bibfnamefont
  {C.}~\bibnamefont {Jeon}}, \bibinfo {author} {\bibfnamefont {F.}~\bibnamefont
  {Sheikh}}, \bibinfo {author} {\bibfnamefont {E.}~\bibnamefont
  {Juarez-Hernandez}}, \bibinfo {author} {\bibfnamefont {B.~P.}\ \bibnamefont
  {Esparza}}, \bibinfo {author} {\bibfnamefont {H.}~\bibnamefont
  {Rampurawala}}, \bibinfo {author} {\bibfnamefont {B.}~\bibnamefont
  {Carlton}}, \bibinfo {author} {\bibfnamefont {S.}~\bibnamefont {Ravikumar}},
  \bibinfo {author} {\bibfnamefont {C.}~\bibnamefont {Nieva}}, \bibinfo
  {author} {\bibfnamefont {S.}~\bibnamefont {Kim}}, \bibinfo {author}
  {\bibfnamefont {H.-J.}\ \bibnamefont {Lee}}, \bibinfo {author} {\bibfnamefont
  {A.}~\bibnamefont {Sammak}}, \bibinfo {author} {\bibfnamefont
  {G.}~\bibnamefont {Scappucci}}, \bibinfo {author} {\bibfnamefont
  {M.}~\bibnamefont {Veldhorst}}, \bibinfo {author} {\bibfnamefont
  {F.}~\bibnamefont {Sebastiano}}, \bibinfo {author} {\bibfnamefont
  {M.}~\bibnamefont {Babaie}}, \bibinfo {author} {\bibfnamefont
  {S.}~\bibnamefont {Pellerano}}, \bibinfo {author} {\bibfnamefont
  {E.}~\bibnamefont {Charbon}},\ and\ \bibinfo {author} {\bibfnamefont
  {L.~M.~K.}\ \bibnamefont {Vandersypen}},\ }\bibfield  {title} {\bibinfo
  {title} {Cmos-based cryogenic control of silicon quantum circuits},\ }\href
  {https://doi.org/10.1038/s41586-021-03469-4} {\bibfield  {journal} {\bibinfo
  {journal} {Nature}\ }\textbf {\bibinfo {volume} {593}},\ \bibinfo {pages}
  {205} (\bibinfo {year} {2021})}\BibitemShut {NoStop}%
\bibitem [{\citenamefont {Pauka}\ \emph {et~al.}(2021)\citenamefont {Pauka},
  \citenamefont {Das}, \citenamefont {Kalra}, \citenamefont {Moini},
  \citenamefont {Yang}, \citenamefont {Trainer}, \citenamefont {Bousquet},
  \citenamefont {Cantaloube}, \citenamefont {Dick}, \citenamefont {Gardner},
  \citenamefont {Manfra},\ and\ \citenamefont
  {Reilly}}]{Pauka2021CMOSControlChip}%
  \BibitemOpen
  \bibfield  {author} {\bibinfo {author} {\bibfnamefont {S.~J.}\ \bibnamefont
  {Pauka}}, \bibinfo {author} {\bibfnamefont {K.}~\bibnamefont {Das}}, \bibinfo
  {author} {\bibfnamefont {R.}~\bibnamefont {Kalra}}, \bibinfo {author}
  {\bibfnamefont {A.}~\bibnamefont {Moini}}, \bibinfo {author} {\bibfnamefont
  {Y.}~\bibnamefont {Yang}}, \bibinfo {author} {\bibfnamefont {M.}~\bibnamefont
  {Trainer}}, \bibinfo {author} {\bibfnamefont {A.}~\bibnamefont {Bousquet}},
  \bibinfo {author} {\bibfnamefont {C.}~\bibnamefont {Cantaloube}}, \bibinfo
  {author} {\bibfnamefont {N.}~\bibnamefont {Dick}}, \bibinfo {author}
  {\bibfnamefont {G.~C.}\ \bibnamefont {Gardner}}, \bibinfo {author}
  {\bibfnamefont {M.~J.}\ \bibnamefont {Manfra}},\ and\ \bibinfo {author}
  {\bibfnamefont {D.~J.}\ \bibnamefont {Reilly}},\ }\bibfield  {title}
  {\bibinfo {title} {A cryogenic cmos chip for generating control signals for
  multiple qubits},\ }\href {https://doi.org/10.1038/s41928-020-00528-y}
  {\bibfield  {journal} {\bibinfo  {journal} {Nature Electronics}\ }\textbf
  {\bibinfo {volume} {4}},\ \bibinfo {pages} {64} (\bibinfo {year}
  {2021})}\BibitemShut {NoStop}%
\bibitem [{\citenamefont {Petit}\ \emph {et~al.}(2020)\citenamefont {Petit},
  \citenamefont {Eenink}, \citenamefont {Russ}, \citenamefont {Lawrie},
  \citenamefont {Hendrickx}, \citenamefont {Philips}, \citenamefont {Clarke},
  \citenamefont {Vandersypen},\ and\ \citenamefont
  {Veldhorst}}]{Petit2020UniversalHotQubits}%
  \BibitemOpen
  \bibfield  {author} {\bibinfo {author} {\bibfnamefont {L.}~\bibnamefont
  {Petit}}, \bibinfo {author} {\bibfnamefont {H.~G.~J.}\ \bibnamefont
  {Eenink}}, \bibinfo {author} {\bibfnamefont {M.}~\bibnamefont {Russ}},
  \bibinfo {author} {\bibfnamefont {W.~I.~L.}\ \bibnamefont {Lawrie}}, \bibinfo
  {author} {\bibfnamefont {N.~W.}\ \bibnamefont {Hendrickx}}, \bibinfo {author}
  {\bibfnamefont {S.~G.~J.}\ \bibnamefont {Philips}}, \bibinfo {author}
  {\bibfnamefont {J.~S.}\ \bibnamefont {Clarke}}, \bibinfo {author}
  {\bibfnamefont {L.~M.~K.}\ \bibnamefont {Vandersypen}},\ and\ \bibinfo
  {author} {\bibfnamefont {M.}~\bibnamefont {Veldhorst}},\ }\bibfield  {title}
  {\bibinfo {title} {Universal quantum logic in hot silicon qubits},\ }\href
  {https://doi.org/10.1038/s41586-020-2170-7} {\bibfield  {journal} {\bibinfo
  {journal} {Nature}\ }\textbf {\bibinfo {volume} {580}},\ \bibinfo {pages}
  {355} (\bibinfo {year} {2020})}\BibitemShut {NoStop}%
\bibitem [{\citenamefont {Yang}\ \emph {et~al.}(2020)\citenamefont {Yang},
  \citenamefont {Leon}, \citenamefont {Hwang}, \citenamefont {Saraiva},
  \citenamefont {Tanttu}, \citenamefont {Huang}, \citenamefont
  {Camirand~Lemyre}, \citenamefont {Chan}, \citenamefont {Tan}, \citenamefont
  {Hudson}, \citenamefont {Itoh}, \citenamefont {Morello}, \citenamefont
  {Pioro-Ladri{\`e}re}, \citenamefont {Laucht},\ and\ \citenamefont
  {Dzurak}}]{Yang2020processor1kelvin}%
  \BibitemOpen
  \bibfield  {author} {\bibinfo {author} {\bibfnamefont {C.~H.}\ \bibnamefont
  {Yang}}, \bibinfo {author} {\bibfnamefont {R.~C.~C.}\ \bibnamefont {Leon}},
  \bibinfo {author} {\bibfnamefont {J.~C.~C.}\ \bibnamefont {Hwang}}, \bibinfo
  {author} {\bibfnamefont {A.}~\bibnamefont {Saraiva}}, \bibinfo {author}
  {\bibfnamefont {T.}~\bibnamefont {Tanttu}}, \bibinfo {author} {\bibfnamefont
  {W.}~\bibnamefont {Huang}}, \bibinfo {author} {\bibfnamefont
  {J.}~\bibnamefont {Camirand~Lemyre}}, \bibinfo {author} {\bibfnamefont
  {K.~W.}\ \bibnamefont {Chan}}, \bibinfo {author} {\bibfnamefont {K.~Y.}\
  \bibnamefont {Tan}}, \bibinfo {author} {\bibfnamefont {F.~E.}\ \bibnamefont
  {Hudson}}, \bibinfo {author} {\bibfnamefont {K.~M.}\ \bibnamefont {Itoh}},
  \bibinfo {author} {\bibfnamefont {A.}~\bibnamefont {Morello}}, \bibinfo
  {author} {\bibfnamefont {M.}~\bibnamefont {Pioro-Ladri{\`e}re}}, \bibinfo
  {author} {\bibfnamefont {A.}~\bibnamefont {Laucht}},\ and\ \bibinfo {author}
  {\bibfnamefont {A.~S.}\ \bibnamefont {Dzurak}},\ }\bibfield  {title}
  {\bibinfo {title} {Operation of a silicon quantum processor unit cell above
  one kelvin},\ }\href {https://doi.org/10.1038/s41586-020-2171-6} {\bibfield
  {journal} {\bibinfo  {journal} {Nature}\ }\textbf {\bibinfo {volume} {580}},\
  \bibinfo {pages} {350} (\bibinfo {year} {2020})}\BibitemShut {NoStop}%
\bibitem [{\citenamefont {Mills}\ \emph {et~al.}(2019)\citenamefont {Mills},
  \citenamefont {Zajac}, \citenamefont {Gullans}, \citenamefont {Schupp},
  \citenamefont {Hazard},\ and\ \citenamefont {Petta}}]{Mills2019}%
  \BibitemOpen
  \bibfield  {author} {\bibinfo {author} {\bibfnamefont {A.}~\bibnamefont
  {Mills}}, \bibinfo {author} {\bibfnamefont {D.}~\bibnamefont {Zajac}},
  \bibinfo {author} {\bibfnamefont {M.}~\bibnamefont {Gullans}}, \bibinfo
  {author} {\bibfnamefont {F.}~\bibnamefont {Schupp}}, \bibinfo {author}
  {\bibfnamefont {T.}~\bibnamefont {Hazard}},\ and\ \bibinfo {author}
  {\bibfnamefont {J.}~\bibnamefont {Petta}},\ }\bibfield  {title} {\bibinfo
  {title} {Shuttling a single charge across a one-dimensional array of silicon
  quantum dots},\ }\href {https://doi.org/10.1038/s41467-019-08970-z}
  {\bibfield  {journal} {\bibinfo  {journal} {Nature communications}\ }\textbf
  {\bibinfo {volume} {10}},\ \bibinfo {pages} {1} (\bibinfo {year}
  {2019})}\BibitemShut {NoStop}%
\bibitem [{\citenamefont {Seidler}\ \emph {et~al.}(2021)\citenamefont
  {Seidler}, \citenamefont {Struck}, \citenamefont {Xue}, \citenamefont
  {Focke}, \citenamefont {Trellenkamp}, \citenamefont {Bluhm},\ and\
  \citenamefont {Schreiber}}]{ConveyorInga}%
  \BibitemOpen
  \bibfield  {author} {\bibinfo {author} {\bibfnamefont {I.}~\bibnamefont
  {Seidler}}, \bibinfo {author} {\bibfnamefont {T.}~\bibnamefont {Struck}},
  \bibinfo {author} {\bibfnamefont {R.}~\bibnamefont {Xue}}, \bibinfo {author}
  {\bibfnamefont {N.}~\bibnamefont {Focke}}, \bibinfo {author} {\bibfnamefont
  {S.}~\bibnamefont {Trellenkamp}}, \bibinfo {author} {\bibfnamefont
  {H.}~\bibnamefont {Bluhm}},\ and\ \bibinfo {author} {\bibfnamefont {L.~R.}\
  \bibnamefont {Schreiber}},\ }\href
  {https://doi.org/10.48550/ARXIV.2108.00879} {\bibinfo {title} {Conveyor-mode
  single-electron shuttling in si/sige for a scalable quantum computing
  architecture}} (\bibinfo {year} {2021}),\ \Eprint
  {https://arxiv.org/abs/2108.00879} {arXiv:2108.00879} \BibitemShut {NoStop}%
\bibitem [{\citenamefont {Noiri}\ \emph
  {et~al.}(2022{\natexlab{b}})\citenamefont {Noiri}, \citenamefont {Takeda},
  \citenamefont {Nakajima}, \citenamefont {Kobayashi}, \citenamefont {Sammak},
  \citenamefont {Scappucci},\ and\ \citenamefont {Tarucha}}]{Shuttling1dot}%
  \BibitemOpen
  \bibfield  {author} {\bibinfo {author} {\bibfnamefont {A.}~\bibnamefont
  {Noiri}}, \bibinfo {author} {\bibfnamefont {K.}~\bibnamefont {Takeda}},
  \bibinfo {author} {\bibfnamefont {T.}~\bibnamefont {Nakajima}}, \bibinfo
  {author} {\bibfnamefont {T.}~\bibnamefont {Kobayashi}}, \bibinfo {author}
  {\bibfnamefont {A.}~\bibnamefont {Sammak}}, \bibinfo {author} {\bibfnamefont
  {G.}~\bibnamefont {Scappucci}},\ and\ \bibinfo {author} {\bibfnamefont
  {S.}~\bibnamefont {Tarucha}},\ }\href
  {https://doi.org/10.48550/ARXIV.2202.01357} {\bibinfo {title} {A
  shuttling-based two-qubit logic gate for linking distant silicon quantum
  processors}} (\bibinfo {year} {2022}{\natexlab{b}}),\ \Eprint
  {https://arxiv.org/abs/2202.01357} {arXiv:2202.01357} \BibitemShut {NoStop}%
\bibitem [{\citenamefont {Nowack}\ \emph {et~al.}(2007)\citenamefont {Nowack},
  \citenamefont {Koppens}, \citenamefont {Nazarov},\ and\ \citenamefont
  {Vandersypen}}]{EDSRSpinOrbit}%
  \BibitemOpen
  \bibfield  {author} {\bibinfo {author} {\bibfnamefont {K.~C.}\ \bibnamefont
  {Nowack}}, \bibinfo {author} {\bibfnamefont {F.~H.~L.}\ \bibnamefont
  {Koppens}}, \bibinfo {author} {\bibfnamefont {Y.~V.}\ \bibnamefont
  {Nazarov}},\ and\ \bibinfo {author} {\bibfnamefont {L.~M.~K.}\ \bibnamefont
  {Vandersypen}},\ }\bibfield  {title} {\bibinfo {title} {Coherent control of a
  single electron spin with electric fields},\ }\href
  {https://doi.org/10.1126/science.1148092} {\bibfield  {journal} {\bibinfo
  {journal} {Science}\ }\textbf {\bibinfo {volume} {318}},\ \bibinfo {pages}
  {1430} (\bibinfo {year} {2007})},\ \Eprint
  {https://arxiv.org/abs/https://www.science.org/doi/pdf/10.1126/science.1148092}
  {https://www.science.org/doi/pdf/10.1126/science.1148092} \BibitemShut
  {NoStop}%
\bibitem [{\citenamefont {Pioro-Ladrière}\ \emph {et~al.}(2008)\citenamefont
  {Pioro-Ladrière}, \citenamefont {Obata}, \citenamefont {Tokura},
  \citenamefont {Shin}, \citenamefont {Kubo}, \citenamefont {Yoshida},
  \citenamefont {Taniyama},\ and\ \citenamefont {Tarucha}}]{EDSRMicroMagnet}%
  \BibitemOpen
  \bibfield  {author} {\bibinfo {author} {\bibfnamefont {M.}~\bibnamefont
  {Pioro-Ladrière}}, \bibinfo {author} {\bibfnamefont {T.}~\bibnamefont
  {Obata}}, \bibinfo {author} {\bibfnamefont {Y.}~\bibnamefont {Tokura}},
  \bibinfo {author} {\bibfnamefont {Y.-S.}\ \bibnamefont {Shin}}, \bibinfo
  {author} {\bibfnamefont {T.}~\bibnamefont {Kubo}}, \bibinfo {author}
  {\bibfnamefont {K.}~\bibnamefont {Yoshida}}, \bibinfo {author} {\bibfnamefont
  {T.}~\bibnamefont {Taniyama}},\ and\ \bibinfo {author} {\bibfnamefont
  {S.}~\bibnamefont {Tarucha}},\ }\bibfield  {title} {\bibinfo {title}
  {{Selective Manipulation of Electron Spins with Electric Fields}},\ }\href
  {https://doi.org/10.1143/PTPS.176.322} {\bibfield  {journal} {\bibinfo
  {journal} {Progress of Theoretical Physics Supplement}\ }\textbf {\bibinfo
  {volume} {176}},\ \bibinfo {pages} {322} (\bibinfo {year} {2008})},\ \Eprint
  {https://arxiv.org/abs/https://doi.org/10.1143/PTPS.176.322}
  {https://doi.org/10.1143/PTPS.176.322} \BibitemShut {NoStop}%
\bibitem [{\citenamefont {Yoneda}\ \emph {et~al.}(2018)\citenamefont {Yoneda},
  \citenamefont {Takeda}, \citenamefont {Otsuka}, \citenamefont {Nakajima},
  \citenamefont {Delbecq}, \citenamefont {Allison}, \citenamefont {Honda},
  \citenamefont {Kodera}, \citenamefont {Oda}, \citenamefont {Hoshi},
  \citenamefont {Usami}, \citenamefont {Itoh},\ and\ \citenamefont
  {Tarucha}}]{Yoneda2018HighFidGates}%
  \BibitemOpen
  \bibfield  {author} {\bibinfo {author} {\bibfnamefont {J.}~\bibnamefont
  {Yoneda}}, \bibinfo {author} {\bibfnamefont {K.}~\bibnamefont {Takeda}},
  \bibinfo {author} {\bibfnamefont {T.}~\bibnamefont {Otsuka}}, \bibinfo
  {author} {\bibfnamefont {T.}~\bibnamefont {Nakajima}}, \bibinfo {author}
  {\bibfnamefont {M.~R.}\ \bibnamefont {Delbecq}}, \bibinfo {author}
  {\bibfnamefont {G.}~\bibnamefont {Allison}}, \bibinfo {author} {\bibfnamefont
  {T.}~\bibnamefont {Honda}}, \bibinfo {author} {\bibfnamefont
  {T.}~\bibnamefont {Kodera}}, \bibinfo {author} {\bibfnamefont
  {S.}~\bibnamefont {Oda}}, \bibinfo {author} {\bibfnamefont {Y.}~\bibnamefont
  {Hoshi}}, \bibinfo {author} {\bibfnamefont {N.}~\bibnamefont {Usami}},
  \bibinfo {author} {\bibfnamefont {K.~M.}\ \bibnamefont {Itoh}},\ and\
  \bibinfo {author} {\bibfnamefont {S.}~\bibnamefont {Tarucha}},\ }\bibfield
  {title} {\bibinfo {title} {A quantum-dot spin qubit with coherence limited by
  charge noise and fidelity higher than 99.9{\%}},\ }\href
  {https://doi.org/10.1038/s41565-017-0014-x} {\bibfield  {journal} {\bibinfo
  {journal} {Nature Nanotechnology}\ }\textbf {\bibinfo {volume} {13}},\
  \bibinfo {pages} {102} (\bibinfo {year} {2018})}\BibitemShut {NoStop}%
\bibitem [{\citenamefont {Struck}\ \emph {et~al.}(2020)\citenamefont {Struck},
  \citenamefont {Hollmann}, \citenamefont {Schauer}, \citenamefont {Fedorets},
  \citenamefont {Schmidbauer}, \citenamefont {Sawano}, \citenamefont {Riemann},
  \citenamefont {Abrosimov}, \citenamefont {Cywi{\'{n}}ski}, \citenamefont
  {Bougeard},\ and\ \citenamefont {Schreiber}}]{Struck2020}%
  \BibitemOpen
  \bibfield  {author} {\bibinfo {author} {\bibfnamefont {T.}~\bibnamefont
  {Struck}}, \bibinfo {author} {\bibfnamefont {A.}~\bibnamefont {Hollmann}},
  \bibinfo {author} {\bibfnamefont {F.}~\bibnamefont {Schauer}}, \bibinfo
  {author} {\bibfnamefont {O.}~\bibnamefont {Fedorets}}, \bibinfo {author}
  {\bibfnamefont {A.}~\bibnamefont {Schmidbauer}}, \bibinfo {author}
  {\bibfnamefont {K.}~\bibnamefont {Sawano}}, \bibinfo {author} {\bibfnamefont
  {H.}~\bibnamefont {Riemann}}, \bibinfo {author} {\bibfnamefont {N.~V.}\
  \bibnamefont {Abrosimov}}, \bibinfo {author} {\bibfnamefont
  {{\L}.}~\bibnamefont {Cywi{\'{n}}ski}}, \bibinfo {author} {\bibfnamefont
  {D.}~\bibnamefont {Bougeard}},\ and\ \bibinfo {author} {\bibfnamefont
  {L.~R.}\ \bibnamefont {Schreiber}},\ }\bibfield  {title} {\bibinfo {title}
  {Low-frequency spin qubit energy splitting noise in highly purified
  28si/sige},\ }\href {https://doi.org/10.1038/s41534-020-0276-2} {\bibfield
  {journal} {\bibinfo  {journal} {npj Quantum Information}\ }\textbf {\bibinfo
  {volume} {6}},\ \bibinfo {pages} {40} (\bibinfo {year} {2020})}\BibitemShut
  {NoStop}%
\bibitem [{\citenamefont {Neumann}\ and\ \citenamefont
  {Schreiber}(2015)}]{NeumannMicroMagnet}%
  \BibitemOpen
  \bibfield  {author} {\bibinfo {author} {\bibfnamefont {R.}~\bibnamefont
  {Neumann}}\ and\ \bibinfo {author} {\bibfnamefont {L.~R.}\ \bibnamefont
  {Schreiber}},\ }\bibfield  {title} {\bibinfo {title} {Simulation of
  micro-magnet stray-field dynamics for spin qubit manipulation},\ }\href
  {https://doi.org/10.1063/1.4921291} {\bibfield  {journal} {\bibinfo
  {journal} {Journal of Applied Physics}\ }\textbf {\bibinfo {volume} {117}},\
  \bibinfo {pages} {193903} (\bibinfo {year} {2015})},\ \Eprint
  {https://arxiv.org/abs/https://doi.org/10.1063/1.4921291}
  {https://doi.org/10.1063/1.4921291} \BibitemShut {NoStop}%
\bibitem [{\citenamefont {Croot}\ \emph {et~al.}(2020)\citenamefont {Croot},
  \citenamefont {Mi}, \citenamefont {Putz}, \citenamefont {Benito},
  \citenamefont {Borjans}, \citenamefont {Burkard},\ and\ \citenamefont
  {Petta}}]{crootflopping}%
  \BibitemOpen
  \bibfield  {author} {\bibinfo {author} {\bibfnamefont {X.}~\bibnamefont
  {Croot}}, \bibinfo {author} {\bibfnamefont {X.}~\bibnamefont {Mi}}, \bibinfo
  {author} {\bibfnamefont {S.}~\bibnamefont {Putz}}, \bibinfo {author}
  {\bibfnamefont {M.}~\bibnamefont {Benito}}, \bibinfo {author} {\bibfnamefont
  {F.}~\bibnamefont {Borjans}}, \bibinfo {author} {\bibfnamefont
  {G.}~\bibnamefont {Burkard}},\ and\ \bibinfo {author} {\bibfnamefont {J.~R.}\
  \bibnamefont {Petta}},\ }\bibfield  {title} {\bibinfo {title} {Flopping-mode
  electric dipole spin resonance},\ }\href
  {https://doi.org/10.1103/PhysRevResearch.2.012006} {\bibfield  {journal}
  {\bibinfo  {journal} {Phys. Rev. Research}\ }\textbf {\bibinfo {volume}
  {2}},\ \bibinfo {pages} {012006} (\bibinfo {year} {2020})}\BibitemShut
  {NoStop}%
\bibitem [{\citenamefont {Benito}\ \emph
  {et~al.}(2019{\natexlab{a}})\citenamefont {Benito}, \citenamefont {Croot},
  \citenamefont {Adelsberger}, \citenamefont {Putz}, \citenamefont {Mi},
  \citenamefont {Petta},\ and\ \citenamefont {Burkard}}]{benitoflopping}%
  \BibitemOpen
  \bibfield  {author} {\bibinfo {author} {\bibfnamefont {M.}~\bibnamefont
  {Benito}}, \bibinfo {author} {\bibfnamefont {X.}~\bibnamefont {Croot}},
  \bibinfo {author} {\bibfnamefont {C.}~\bibnamefont {Adelsberger}}, \bibinfo
  {author} {\bibfnamefont {S.}~\bibnamefont {Putz}}, \bibinfo {author}
  {\bibfnamefont {X.}~\bibnamefont {Mi}}, \bibinfo {author} {\bibfnamefont
  {J.~R.}\ \bibnamefont {Petta}},\ and\ \bibinfo {author} {\bibfnamefont
  {G.}~\bibnamefont {Burkard}},\ }\bibfield  {title} {\bibinfo {title}
  {Electric-field control and noise protection of the flopping-mode spin
  qubit},\ }\href {https://doi.org/10.1103/PhysRevB.100.125430} {\bibfield
  {journal} {\bibinfo  {journal} {Phys. Rev. B}\ }\textbf {\bibinfo {volume}
  {100}},\ \bibinfo {pages} {125430} (\bibinfo {year}
  {2019}{\natexlab{a}})}\BibitemShut {NoStop}%
\bibitem [{\citenamefont {Huang}\ and\ \citenamefont
  {Hu}(2014)}]{HuSpinHotspot}%
  \BibitemOpen
  \bibfield  {author} {\bibinfo {author} {\bibfnamefont {P.}~\bibnamefont
  {Huang}}\ and\ \bibinfo {author} {\bibfnamefont {X.}~\bibnamefont {Hu}},\
  }\bibfield  {title} {\bibinfo {title} {Spin relaxation in a si quantum dot
  due to spin-valley mixing},\ }\href
  {https://doi.org/10.1103/PhysRevB.90.235315} {\bibfield  {journal} {\bibinfo
  {journal} {Phys. Rev. B}\ }\textbf {\bibinfo {volume} {90}},\ \bibinfo
  {pages} {235315} (\bibinfo {year} {2014})}\BibitemShut {NoStop}%
\bibitem [{\citenamefont {Benito}\ \emph {et~al.}(2017)\citenamefont {Benito},
  \citenamefont {Mi}, \citenamefont {Taylor}, \citenamefont {Petta},\ and\
  \citenamefont {Burkard}}]{Input-outputBenito}%
  \BibitemOpen
  \bibfield  {author} {\bibinfo {author} {\bibfnamefont {M.}~\bibnamefont
  {Benito}}, \bibinfo {author} {\bibfnamefont {X.}~\bibnamefont {Mi}}, \bibinfo
  {author} {\bibfnamefont {J.~M.}\ \bibnamefont {Taylor}}, \bibinfo {author}
  {\bibfnamefont {J.~R.}\ \bibnamefont {Petta}},\ and\ \bibinfo {author}
  {\bibfnamefont {G.}~\bibnamefont {Burkard}},\ }\bibfield  {title} {\bibinfo
  {title} {Input-output theory for spin-photon coupling in si double quantum
  dots},\ }\href {https://doi.org/10.1103/PhysRevB.96.235434} {\bibfield
  {journal} {\bibinfo  {journal} {Phys. Rev. B}\ }\textbf {\bibinfo {volume}
  {96}},\ \bibinfo {pages} {235434} (\bibinfo {year} {2017})}\BibitemShut
  {NoStop}%
\bibitem [{\citenamefont {Mi}\ \emph {et~al.}(2018)\citenamefont {Mi},
  \citenamefont {Benito}, \citenamefont {Putz}, \citenamefont {Zajac},
  \citenamefont {Taylor}, \citenamefont {Burkard},\ and\ \citenamefont
  {Petta}}]{Mi2018FloppingSpinPhotonInterface}%
  \BibitemOpen
  \bibfield  {author} {\bibinfo {author} {\bibfnamefont {X.}~\bibnamefont
  {Mi}}, \bibinfo {author} {\bibfnamefont {M.}~\bibnamefont {Benito}}, \bibinfo
  {author} {\bibfnamefont {S.}~\bibnamefont {Putz}}, \bibinfo {author}
  {\bibfnamefont {D.~M.}\ \bibnamefont {Zajac}}, \bibinfo {author}
  {\bibfnamefont {J.~M.}\ \bibnamefont {Taylor}}, \bibinfo {author}
  {\bibfnamefont {G.}~\bibnamefont {Burkard}},\ and\ \bibinfo {author}
  {\bibfnamefont {J.~R.}\ \bibnamefont {Petta}},\ }\bibfield  {title} {\bibinfo
  {title} {A coherent spin--photon interface in silicon},\ }\href
  {https://doi.org/10.1038/nature25769} {\bibfield  {journal} {\bibinfo
  {journal} {Nature}\ }\textbf {\bibinfo {volume} {555}},\ \bibinfo {pages}
  {599} (\bibinfo {year} {2018})}\BibitemShut {NoStop}%
\bibitem [{\citenamefont {Samkharadze}\ \emph {et~al.}(2018)\citenamefont
  {Samkharadze}, \citenamefont {Zheng}, \citenamefont {Kalhor}, \citenamefont
  {Brousse}, \citenamefont {Sammak}, \citenamefont {Mendes}, \citenamefont
  {Blais}, \citenamefont {Scappucci},\ and\ \citenamefont
  {Vandersypen}}]{FloppingSpinPhotonCouplingSi}%
  \BibitemOpen
  \bibfield  {author} {\bibinfo {author} {\bibfnamefont {N.}~\bibnamefont
  {Samkharadze}}, \bibinfo {author} {\bibfnamefont {G.}~\bibnamefont {Zheng}},
  \bibinfo {author} {\bibfnamefont {N.}~\bibnamefont {Kalhor}}, \bibinfo
  {author} {\bibfnamefont {D.}~\bibnamefont {Brousse}}, \bibinfo {author}
  {\bibfnamefont {A.}~\bibnamefont {Sammak}}, \bibinfo {author} {\bibfnamefont
  {U.~C.}\ \bibnamefont {Mendes}}, \bibinfo {author} {\bibfnamefont
  {A.}~\bibnamefont {Blais}}, \bibinfo {author} {\bibfnamefont
  {G.}~\bibnamefont {Scappucci}},\ and\ \bibinfo {author} {\bibfnamefont
  {L.~M.~K.}\ \bibnamefont {Vandersypen}},\ }\bibfield  {title} {\bibinfo
  {title} {Strong spin-photon coupling in silicon},\ }\href
  {https://doi.org/10.1126/science.aar4054} {\bibfield  {journal} {\bibinfo
  {journal} {Science}\ }\textbf {\bibinfo {volume} {359}},\ \bibinfo {pages}
  {1123} (\bibinfo {year} {2018})},\ \Eprint
  {https://arxiv.org/abs/https://www.science.org/doi/pdf/10.1126/science.aar4054}
  {https://www.science.org/doi/pdf/10.1126/science.aar4054} \BibitemShut
  {NoStop}%
\bibitem [{\citenamefont {Mi}\ \emph {et~al.}(2017)\citenamefont {Mi},
  \citenamefont {Cady}, \citenamefont {Zajac}, \citenamefont {Deelman},\ and\
  \citenamefont {Petta}}]{CouplingSingleElectronMW}%
  \BibitemOpen
  \bibfield  {author} {\bibinfo {author} {\bibfnamefont {X.}~\bibnamefont
  {Mi}}, \bibinfo {author} {\bibfnamefont {J.~V.}\ \bibnamefont {Cady}},
  \bibinfo {author} {\bibfnamefont {D.~M.}\ \bibnamefont {Zajac}}, \bibinfo
  {author} {\bibfnamefont {P.~W.}\ \bibnamefont {Deelman}},\ and\ \bibinfo
  {author} {\bibfnamefont {J.~R.}\ \bibnamefont {Petta}},\ }\bibfield  {title}
  {\bibinfo {title} {Strong coupling of a single electron in silicon to a
  microwave photon},\ }\href {https://doi.org/10.1126/science.aal2469}
  {\bibfield  {journal} {\bibinfo  {journal} {Science}\ }\textbf {\bibinfo
  {volume} {355}},\ \bibinfo {pages} {156} (\bibinfo {year} {2017})},\ \Eprint
  {https://arxiv.org/abs/https://www.science.org/doi/pdf/10.1126/science.aal2469}
  {https://www.science.org/doi/pdf/10.1126/science.aal2469} \BibitemShut
  {NoStop}%
\bibitem [{\citenamefont {Stockklauser}\ \emph {et~al.}(2017)\citenamefont
  {Stockklauser}, \citenamefont {Scarlino}, \citenamefont {Koski},
  \citenamefont {Gasparinetti}, \citenamefont {Andersen}, \citenamefont
  {Reichl}, \citenamefont {Wegscheider}, \citenamefont {Ihn}, \citenamefont
  {Ensslin},\ and\ \citenamefont {Wallraff}}]{StrongCouplingfloppingCavity}%
  \BibitemOpen
  \bibfield  {author} {\bibinfo {author} {\bibfnamefont {A.}~\bibnamefont
  {Stockklauser}}, \bibinfo {author} {\bibfnamefont {P.}~\bibnamefont
  {Scarlino}}, \bibinfo {author} {\bibfnamefont {J.~V.}\ \bibnamefont {Koski}},
  \bibinfo {author} {\bibfnamefont {S.}~\bibnamefont {Gasparinetti}}, \bibinfo
  {author} {\bibfnamefont {C.~K.}\ \bibnamefont {Andersen}}, \bibinfo {author}
  {\bibfnamefont {C.}~\bibnamefont {Reichl}}, \bibinfo {author} {\bibfnamefont
  {W.}~\bibnamefont {Wegscheider}}, \bibinfo {author} {\bibfnamefont
  {T.}~\bibnamefont {Ihn}}, \bibinfo {author} {\bibfnamefont {K.}~\bibnamefont
  {Ensslin}},\ and\ \bibinfo {author} {\bibfnamefont {A.}~\bibnamefont
  {Wallraff}},\ }\bibfield  {title} {\bibinfo {title} {Strong coupling cavity
  qed with gate-defined double quantum dots enabled by a high impedance
  resonator},\ }\href {https://doi.org/10.1103/PhysRevX.7.011030} {\bibfield
  {journal} {\bibinfo  {journal} {Phys. Rev. X}\ }\textbf {\bibinfo {volume}
  {7}},\ \bibinfo {pages} {011030} (\bibinfo {year} {2017})}\BibitemShut
  {NoStop}%
\bibitem [{\citenamefont {Benito}\ \emph
  {et~al.}(2019{\natexlab{b}})\citenamefont {Benito}, \citenamefont {Petta},\
  and\ \citenamefont {Burkard}}]{Flopping2QubitCavityMediated}%
  \BibitemOpen
  \bibfield  {author} {\bibinfo {author} {\bibfnamefont {M.}~\bibnamefont
  {Benito}}, \bibinfo {author} {\bibfnamefont {J.~R.}\ \bibnamefont {Petta}},\
  and\ \bibinfo {author} {\bibfnamefont {G.}~\bibnamefont {Burkard}},\
  }\bibfield  {title} {\bibinfo {title} {Optimized cavity-mediated dispersive
  two-qubit gates between spin qubits},\ }\href
  {https://doi.org/10.1103/PhysRevB.100.081412} {\bibfield  {journal} {\bibinfo
   {journal} {Phys. Rev. B}\ }\textbf {\bibinfo {volume} {100}},\ \bibinfo
  {pages} {081412} (\bibinfo {year} {2019}{\natexlab{b}})}\BibitemShut
  {NoStop}%
\bibitem [{\citenamefont {Cayao}\ \emph {et~al.}(2020)\citenamefont {Cayao},
  \citenamefont {Benito},\ and\ \citenamefont
  {Burkard}}]{flopping2qubitcapacitively}%
  \BibitemOpen
  \bibfield  {author} {\bibinfo {author} {\bibfnamefont {J.}~\bibnamefont
  {Cayao}}, \bibinfo {author} {\bibfnamefont {M.}~\bibnamefont {Benito}},\ and\
  \bibinfo {author} {\bibfnamefont {G.}~\bibnamefont {Burkard}},\ }\bibfield
  {title} {\bibinfo {title} {Programmable two-qubit gates in capacitively
  coupled flopping-mode spin qubits},\ }\href
  {https://doi.org/10.1103/PhysRevB.101.195438} {\bibfield  {journal} {\bibinfo
   {journal} {Phys. Rev. B}\ }\textbf {\bibinfo {volume} {101}},\ \bibinfo
  {pages} {195438} (\bibinfo {year} {2020})}\BibitemShut {NoStop}%
\bibitem [{\citenamefont {Mutter}\ and\ \citenamefont
  {Burkard}(2021)}]{FloppingGeHole}%
  \BibitemOpen
  \bibfield  {author} {\bibinfo {author} {\bibfnamefont {P.~M.}\ \bibnamefont
  {Mutter}}\ and\ \bibinfo {author} {\bibfnamefont {G.}~\bibnamefont
  {Burkard}},\ }\bibfield  {title} {\bibinfo {title} {Natural heavy-hole
  flopping mode qubit in germanium},\ }\href
  {https://doi.org/10.1103/PhysRevResearch.3.013194} {\bibfield  {journal}
  {\bibinfo  {journal} {Phys. Rev. Research}\ }\textbf {\bibinfo {volume}
  {3}},\ \bibinfo {pages} {013194} (\bibinfo {year} {2021})}\BibitemShut
  {NoStop}%
\bibitem [{\citenamefont {Reiher}\ and\ \citenamefont
  {B{\'e}rub{\'e}-Lauzi{\`e}re}(2021)}]{FloppingModeRegimeTransition}%
  \BibitemOpen
  \bibfield  {author} {\bibinfo {author} {\bibfnamefont {V.}~\bibnamefont
  {Reiher}}\ and\ \bibinfo {author} {\bibfnamefont {Y.}~\bibnamefont
  {B{\'e}rub{\'e}-Lauzi{\`e}re}},\ }\bibfield  {title} {\bibinfo {title}
  {Optimal control of the operating regime of a single electron double quantum
  dot},\ }\href {https://arxiv.org/abs/2104.13571} {\bibfield  {journal}
  {\bibinfo  {journal} {arXiv preprint arXiv:2104.13571}\ } (\bibinfo {year}
  {2021})}\BibitemShut {NoStop}%
\bibitem [{\citenamefont {Wuetz}\ \emph {et~al.}(2021)\citenamefont {Wuetz},
  \citenamefont {Losert}, \citenamefont {Koelling}, \citenamefont {Stehouwer},
  \citenamefont {Zwerver}, \citenamefont {Philips}, \citenamefont {Madzik},
  \citenamefont {Xue}, \citenamefont {Zheng}, \citenamefont {Lodari},
  \citenamefont {Amitonov}, \citenamefont {Samkharadze}, \citenamefont
  {Sammak}, \citenamefont {Vandersypen}, \citenamefont {Rahman}, \citenamefont
  {Coppersmith}, \citenamefont {Moutanabbir}, \citenamefont {Friesen},\ and\
  \citenamefont {Scappucci}}]{Gefluctuations}%
  \BibitemOpen
  \bibfield  {author} {\bibinfo {author} {\bibfnamefont {B.~P.}\ \bibnamefont
  {Wuetz}}, \bibinfo {author} {\bibfnamefont {M.~P.}\ \bibnamefont {Losert}},
  \bibinfo {author} {\bibfnamefont {S.}~\bibnamefont {Koelling}}, \bibinfo
  {author} {\bibfnamefont {L.~E.~A.}\ \bibnamefont {Stehouwer}}, \bibinfo
  {author} {\bibfnamefont {A.-M.~J.}\ \bibnamefont {Zwerver}}, \bibinfo
  {author} {\bibfnamefont {S.~G.~J.}\ \bibnamefont {Philips}}, \bibinfo
  {author} {\bibfnamefont {M.~T.}\ \bibnamefont {Madzik}}, \bibinfo {author}
  {\bibfnamefont {X.}~\bibnamefont {Xue}}, \bibinfo {author} {\bibfnamefont
  {G.}~\bibnamefont {Zheng}}, \bibinfo {author} {\bibfnamefont
  {M.}~\bibnamefont {Lodari}}, \bibinfo {author} {\bibfnamefont {S.~V.}\
  \bibnamefont {Amitonov}}, \bibinfo {author} {\bibfnamefont {N.}~\bibnamefont
  {Samkharadze}}, \bibinfo {author} {\bibfnamefont {A.}~\bibnamefont {Sammak}},
  \bibinfo {author} {\bibfnamefont {L.~M.~K.}\ \bibnamefont {Vandersypen}},
  \bibinfo {author} {\bibfnamefont {R.}~\bibnamefont {Rahman}}, \bibinfo
  {author} {\bibfnamefont {S.~N.}\ \bibnamefont {Coppersmith}}, \bibinfo
  {author} {\bibfnamefont {O.}~\bibnamefont {Moutanabbir}}, \bibinfo {author}
  {\bibfnamefont {M.}~\bibnamefont {Friesen}},\ and\ \bibinfo {author}
  {\bibfnamefont {G.}~\bibnamefont {Scappucci}},\ }\href
  {https://doi.org/10.48550/ARXIV.2112.09606} {\bibinfo {title} {Atomic
  fluctuations lifting the energy degeneracy in si/sige quantum dots}}
  (\bibinfo {year} {2021}),\ \Eprint {https://arxiv.org/abs/2112.09606}
  {arXiv:2112.09606} \BibitemShut {NoStop}%
\bibitem [{\citenamefont {Teske}(2020)}]{qopt-applications}%
  \BibitemOpen
  \bibfield  {author} {\bibinfo {author} {\bibfnamefont {J.}~\bibnamefont
  {Teske}},\ }\href@noop {} {\bibinfo {title} {{qopt-applications}: Simulations
  and optimal control implemented with qopt}},\ \bibinfo {howpublished}
  {https://github.com/qutech/qopt-applications} (\bibinfo {year}
  {2020})\BibitemShut {NoStop}%
\bibitem [{\citenamefont {Teske}\ \emph {et~al.}(2022)\citenamefont {Teske},
  \citenamefont {Cerfontaine},\ and\ \citenamefont {Bluhm}}]{qoptPaper}%
  \BibitemOpen
  \bibfield  {author} {\bibinfo {author} {\bibfnamefont {J.~D.}\ \bibnamefont
  {Teske}}, \bibinfo {author} {\bibfnamefont {P.}~\bibnamefont {Cerfontaine}},\
  and\ \bibinfo {author} {\bibfnamefont {H.}~\bibnamefont {Bluhm}},\ }\bibfield
   {title} {\bibinfo {title} {qopt: An experiment-oriented software package for
  qubit simulation and quantum optimal control},\ }\href
  {https://doi.org/10.1103/PhysRevApplied.17.034036} {\bibfield  {journal}
  {\bibinfo  {journal} {Phys. Rev. Applied}\ }\textbf {\bibinfo {volume}
  {17}},\ \bibinfo {pages} {034036} (\bibinfo {year} {2022})}\BibitemShut
  {NoStop}%
\bibitem [{\citenamefont {Wood}\ and\ \citenamefont
  {Gambetta}(2018)}]{LeakageCostFunction}%
  \BibitemOpen
  \bibfield  {author} {\bibinfo {author} {\bibfnamefont {C.~J.}\ \bibnamefont
  {Wood}}\ and\ \bibinfo {author} {\bibfnamefont {J.~M.}\ \bibnamefont
  {Gambetta}},\ }\bibfield  {title} {\bibinfo {title} {Quantification and
  characterization of leakage errors},\ }\href
  {https://doi.org/10.1103/PhysRevA.97.032306} {\bibfield  {journal} {\bibinfo
  {journal} {Phys. Rev. A}\ }\textbf {\bibinfo {volume} {97}},\ \bibinfo
  {pages} {032306} (\bibinfo {year} {2018})}\BibitemShut {NoStop}%
\bibitem [{\citenamefont {Shkolnikov}\ \emph {et~al.}(2020)\citenamefont
  {Shkolnikov}, \citenamefont {Mauch},\ and\ \citenamefont
  {Burkard}}]{PRBHolonomicAverageGateFid}%
  \BibitemOpen
  \bibfield  {author} {\bibinfo {author} {\bibfnamefont {V.~O.}\ \bibnamefont
  {Shkolnikov}}, \bibinfo {author} {\bibfnamefont {R.}~\bibnamefont {Mauch}},\
  and\ \bibinfo {author} {\bibfnamefont {G.}~\bibnamefont {Burkard}},\
  }\bibfield  {title} {\bibinfo {title} {All-microwave holonomic control of an
  electron-nuclear two-qubit register in diamond},\ }\href
  {https://doi.org/10.1103/PhysRevB.101.155306} {\bibfield  {journal} {\bibinfo
   {journal} {Phys. Rev. B}\ }\textbf {\bibinfo {volume} {101}},\ \bibinfo
  {pages} {155306} (\bibinfo {year} {2020})}\BibitemShut {NoStop}%
\bibitem [{\citenamefont {Dial}\ \emph {et~al.}(2013)\citenamefont {Dial},
  \citenamefont {Shulman}, \citenamefont {Harvey}, \citenamefont {Bluhm},
  \citenamefont {Umansky},\ and\ \citenamefont
  {Yacoby}}]{DialNoiseSpectroscopy}%
  \BibitemOpen
  \bibfield  {author} {\bibinfo {author} {\bibfnamefont {O.~E.}\ \bibnamefont
  {Dial}}, \bibinfo {author} {\bibfnamefont {M.~D.}\ \bibnamefont {Shulman}},
  \bibinfo {author} {\bibfnamefont {S.~P.}\ \bibnamefont {Harvey}}, \bibinfo
  {author} {\bibfnamefont {H.}~\bibnamefont {Bluhm}}, \bibinfo {author}
  {\bibfnamefont {V.}~\bibnamefont {Umansky}},\ and\ \bibinfo {author}
  {\bibfnamefont {A.}~\bibnamefont {Yacoby}},\ }\bibfield  {title} {\bibinfo
  {title} {Charge noise spectroscopy using coherent exchange oscillations in a
  singlet-triplet qubit},\ }\href
  {https://doi.org/10.1103/PhysRevLett.110.146804} {\bibfield  {journal}
  {\bibinfo  {journal} {Phys. Rev. Lett.}\ }\textbf {\bibinfo {volume} {110}},\
  \bibinfo {pages} {146804} (\bibinfo {year} {2013})}\BibitemShut {NoStop}%
\bibitem [{\citenamefont {Kawakami}\ \emph {et~al.}(2014)\citenamefont
  {Kawakami}, \citenamefont {Scarlino}, \citenamefont {Ward}, \citenamefont
  {Braakman}, \citenamefont {Savage}, \citenamefont {Lagally}, \citenamefont
  {Friesen}, \citenamefont {Coppersmith}, \citenamefont {Eriksson},\ and\
  \citenamefont {Vandersypen}}]{HFINaturalSiKawakami2014-ek}%
  \BibitemOpen
  \bibfield  {author} {\bibinfo {author} {\bibfnamefont {E.}~\bibnamefont
  {Kawakami}}, \bibinfo {author} {\bibfnamefont {P.}~\bibnamefont {Scarlino}},
  \bibinfo {author} {\bibfnamefont {D.~R.}\ \bibnamefont {Ward}}, \bibinfo
  {author} {\bibfnamefont {F.~R.}\ \bibnamefont {Braakman}}, \bibinfo {author}
  {\bibfnamefont {D.~E.}\ \bibnamefont {Savage}}, \bibinfo {author}
  {\bibfnamefont {M.~G.}\ \bibnamefont {Lagally}}, \bibinfo {author}
  {\bibfnamefont {M.}~\bibnamefont {Friesen}}, \bibinfo {author} {\bibfnamefont
  {S.~N.}\ \bibnamefont {Coppersmith}}, \bibinfo {author} {\bibfnamefont
  {M.~A.}\ \bibnamefont {Eriksson}},\ and\ \bibinfo {author} {\bibfnamefont
  {L.~M.~K.}\ \bibnamefont {Vandersypen}},\ }\bibfield  {title} {\bibinfo
  {title} {Electrical control of a long-lived spin qubit in a {Si/SiGe} quantum
  dot},\ }\href@noop {} {\bibfield  {journal} {\bibinfo  {journal} {Nature
  Nanotechnology}\ }\textbf {\bibinfo {volume} {9}},\ \bibinfo {pages} {666}
  (\bibinfo {year} {2014})}\BibitemShut {NoStop}%
\bibitem [{\citenamefont {Assali}\ \emph {et~al.}(2011)\citenamefont {Assali},
  \citenamefont {Petrilli}, \citenamefont {Capaz}, \citenamefont {Koiller},
  \citenamefont {Hu},\ and\ \citenamefont
  {Das~Sarma}}]{ConcentrationSi29HyperfineTheoretical}%
  \BibitemOpen
  \bibfield  {author} {\bibinfo {author} {\bibfnamefont {L.~V.~C.}\
  \bibnamefont {Assali}}, \bibinfo {author} {\bibfnamefont {H.~M.}\
  \bibnamefont {Petrilli}}, \bibinfo {author} {\bibfnamefont {R.~B.}\
  \bibnamefont {Capaz}}, \bibinfo {author} {\bibfnamefont {B.}~\bibnamefont
  {Koiller}}, \bibinfo {author} {\bibfnamefont {X.}~\bibnamefont {Hu}},\ and\
  \bibinfo {author} {\bibfnamefont {S.}~\bibnamefont {Das~Sarma}},\ }\bibfield
  {title} {\bibinfo {title} {Hyperfine interactions in silicon quantum dots},\
  }\href {https://doi.org/10.1103/PhysRevB.83.165301} {\bibfield  {journal}
  {\bibinfo  {journal} {Phys. Rev. B}\ }\textbf {\bibinfo {volume} {83}},\
  \bibinfo {pages} {165301} (\bibinfo {year} {2011})}\BibitemShut {NoStop}%
\bibitem [{\citenamefont {Shevchenko}\ \emph {et~al.}(2010)\citenamefont
  {Shevchenko}, \citenamefont {Ashhab},\ and\ \citenamefont
  {Nori}}]{LZS_SHEVCHENKO20101}%
  \BibitemOpen
  \bibfield  {author} {\bibinfo {author} {\bibfnamefont {S.}~\bibnamefont
  {Shevchenko}}, \bibinfo {author} {\bibfnamefont {S.}~\bibnamefont {Ashhab}},\
  and\ \bibinfo {author} {\bibfnamefont {F.}~\bibnamefont {Nori}},\ }\bibfield
  {title} {\bibinfo {title} {Landau–zener–stückelberg interferometry},\
  }\href {https://doi.org/https://doi.org/10.1016/j.physrep.2010.03.002}
  {\bibfield  {journal} {\bibinfo  {journal} {Physics Reports}\ }\textbf
  {\bibinfo {volume} {492}},\ \bibinfo {pages} {1} (\bibinfo {year}
  {2010})}\BibitemShut {NoStop}%
\bibitem [{\citenamefont {Ando}\ \emph {et~al.}(1982)\citenamefont {Ando},
  \citenamefont {Fowler},\ and\ \citenamefont {Stern}}]{ValleyBasicsAndo1982}%
  \BibitemOpen
  \bibfield  {author} {\bibinfo {author} {\bibfnamefont {T.}~\bibnamefont
  {Ando}}, \bibinfo {author} {\bibfnamefont {A.~B.}\ \bibnamefont {Fowler}},\
  and\ \bibinfo {author} {\bibfnamefont {F.}~\bibnamefont {Stern}},\ }\bibfield
   {title} {\bibinfo {title} {Electronic properties of two-dimensional
  systems},\ }\href {https://doi.org/10.1103/RevModPhys.54.437} {\bibfield
  {journal} {\bibinfo  {journal} {Rev. Mod. Phys.}\ }\textbf {\bibinfo {volume}
  {54}},\ \bibinfo {pages} {437} (\bibinfo {year} {1982})}\BibitemShut
  {NoStop}%
\bibitem [{\citenamefont {Saraiva}\ \emph {et~al.}(2009)\citenamefont
  {Saraiva}, \citenamefont {Calder\'on}, \citenamefont {Hu}, \citenamefont
  {Das~Sarma},\ and\ \citenamefont {Koiller}}]{ValleyInterfacePhysics}%
  \BibitemOpen
  \bibfield  {author} {\bibinfo {author} {\bibfnamefont {A.~L.}\ \bibnamefont
  {Saraiva}}, \bibinfo {author} {\bibfnamefont {M.~J.}\ \bibnamefont
  {Calder\'on}}, \bibinfo {author} {\bibfnamefont {X.}~\bibnamefont {Hu}},
  \bibinfo {author} {\bibfnamefont {S.}~\bibnamefont {Das~Sarma}},\ and\
  \bibinfo {author} {\bibfnamefont {B.}~\bibnamefont {Koiller}},\ }\bibfield
  {title} {\bibinfo {title} {Physical mechanisms of interface-mediated
  intervalley coupling in si},\ }\href
  {https://doi.org/10.1103/PhysRevB.80.081305} {\bibfield  {journal} {\bibinfo
  {journal} {Phys. Rev. B}\ }\textbf {\bibinfo {volume} {80}},\ \bibinfo
  {pages} {081305} (\bibinfo {year} {2009})}\BibitemShut {NoStop}%
\bibitem [{\citenamefont {Sham}\ and\ \citenamefont
  {Nakayama}(1979)}]{ValleySplittingEffectiveMass}%
  \BibitemOpen
  \bibfield  {author} {\bibinfo {author} {\bibfnamefont {L.~J.}\ \bibnamefont
  {Sham}}\ and\ \bibinfo {author} {\bibfnamefont {M.}~\bibnamefont
  {Nakayama}},\ }\bibfield  {title} {\bibinfo {title} {Effective-mass
  approximation in the presence of an interface},\ }\href
  {https://doi.org/10.1103/PhysRevB.20.734} {\bibfield  {journal} {\bibinfo
  {journal} {Phys. Rev. B}\ }\textbf {\bibinfo {volume} {20}},\ \bibinfo
  {pages} {734} (\bibinfo {year} {1979})}\BibitemShut {NoStop}%
\bibitem [{\citenamefont {Yang}\ \emph {et~al.}(2013)\citenamefont {Yang},
  \citenamefont {Rossi}, \citenamefont {Ruskov}, \citenamefont {Lai},
  \citenamefont {Mohiyaddin}, \citenamefont {Lee}, \citenamefont {Tahan},
  \citenamefont {Klimeck}, \citenamefont {Morello},\ and\ \citenamefont
  {Dzurak}}]{TunableValleySplitting}%
  \BibitemOpen
  \bibfield  {author} {\bibinfo {author} {\bibfnamefont {C.~H.}\ \bibnamefont
  {Yang}}, \bibinfo {author} {\bibfnamefont {A.}~\bibnamefont {Rossi}},
  \bibinfo {author} {\bibfnamefont {R.}~\bibnamefont {Ruskov}}, \bibinfo
  {author} {\bibfnamefont {N.~S.}\ \bibnamefont {Lai}}, \bibinfo {author}
  {\bibfnamefont {F.~A.}\ \bibnamefont {Mohiyaddin}}, \bibinfo {author}
  {\bibfnamefont {S.}~\bibnamefont {Lee}}, \bibinfo {author} {\bibfnamefont
  {C.}~\bibnamefont {Tahan}}, \bibinfo {author} {\bibfnamefont
  {G.}~\bibnamefont {Klimeck}}, \bibinfo {author} {\bibfnamefont
  {A.}~\bibnamefont {Morello}},\ and\ \bibinfo {author} {\bibfnamefont {A.~S.}\
  \bibnamefont {Dzurak}},\ }\bibfield  {title} {\bibinfo {title} {Spin-valley
  lifetimes in a silicon quantum dot with tunable valley splitting},\ }\href
  {https://doi.org/10.1038/ncomms3069} {\bibfield  {journal} {\bibinfo
  {journal} {Nature Communications}\ }\textbf {\bibinfo {volume} {4}},\
  \bibinfo {pages} {2069} (\bibinfo {year} {2013})}\BibitemShut {NoStop}%
\bibitem [{\citenamefont {McJunkin}\ \emph
  {et~al.}(2021{\natexlab{a}})\citenamefont {McJunkin}, \citenamefont
  {MacQuarrie}, \citenamefont {Tom}, \citenamefont {Neyens}, \citenamefont
  {Dodson}, \citenamefont {Thorgrimsson}, \citenamefont {Corrigan},
  \citenamefont {Ercan}, \citenamefont {Savage}, \citenamefont {Lagally},
  \citenamefont {Joynt}, \citenamefont {Coppersmith}, \citenamefont {Friesen},\
  and\ \citenamefont {Eriksson}}]{GeSpikeValley}%
  \BibitemOpen
  \bibfield  {author} {\bibinfo {author} {\bibfnamefont {T.}~\bibnamefont
  {McJunkin}}, \bibinfo {author} {\bibfnamefont {E.~R.}\ \bibnamefont
  {MacQuarrie}}, \bibinfo {author} {\bibfnamefont {L.}~\bibnamefont {Tom}},
  \bibinfo {author} {\bibfnamefont {S.~F.}\ \bibnamefont {Neyens}}, \bibinfo
  {author} {\bibfnamefont {J.~P.}\ \bibnamefont {Dodson}}, \bibinfo {author}
  {\bibfnamefont {B.}~\bibnamefont {Thorgrimsson}}, \bibinfo {author}
  {\bibfnamefont {J.}~\bibnamefont {Corrigan}}, \bibinfo {author}
  {\bibfnamefont {H.~E.}\ \bibnamefont {Ercan}}, \bibinfo {author}
  {\bibfnamefont {D.~E.}\ \bibnamefont {Savage}}, \bibinfo {author}
  {\bibfnamefont {M.~G.}\ \bibnamefont {Lagally}}, \bibinfo {author}
  {\bibfnamefont {R.}~\bibnamefont {Joynt}}, \bibinfo {author} {\bibfnamefont
  {S.~N.}\ \bibnamefont {Coppersmith}}, \bibinfo {author} {\bibfnamefont
  {M.}~\bibnamefont {Friesen}},\ and\ \bibinfo {author} {\bibfnamefont {M.~A.}\
  \bibnamefont {Eriksson}},\ }\bibfield  {title} {\bibinfo {title} {Valley
  splittings in si/sige quantum dots with a germanium spike in the silicon
  well},\ }\href {https://doi.org/10.1103/PhysRevB.104.085406} {\bibfield
  {journal} {\bibinfo  {journal} {Phys. Rev. B}\ }\textbf {\bibinfo {volume}
  {104}},\ \bibinfo {pages} {085406} (\bibinfo {year}
  {2021}{\natexlab{a}})}\BibitemShut {NoStop}%
\bibitem [{\citenamefont {McJunkin}\ \emph
  {et~al.}(2021{\natexlab{b}})\citenamefont {McJunkin}, \citenamefont {Harpt},
  \citenamefont {Feng}, \citenamefont {Losert}, \citenamefont {Rahman},
  \citenamefont {Dodson}, \citenamefont {Wolfe}, \citenamefont {Savage},
  \citenamefont {Lagally}, \citenamefont {Coppersmith}, \citenamefont
  {Friesen}, \citenamefont {Joynt},\ and\ \citenamefont
  {Eriksson}}]{GeOscillations}%
  \BibitemOpen
  \bibfield  {author} {\bibinfo {author} {\bibfnamefont {T.}~\bibnamefont
  {McJunkin}}, \bibinfo {author} {\bibfnamefont {B.}~\bibnamefont {Harpt}},
  \bibinfo {author} {\bibfnamefont {Y.}~\bibnamefont {Feng}}, \bibinfo {author}
  {\bibfnamefont {M.}~\bibnamefont {Losert}}, \bibinfo {author} {\bibfnamefont
  {R.}~\bibnamefont {Rahman}}, \bibinfo {author} {\bibfnamefont {J.~P.}\
  \bibnamefont {Dodson}}, \bibinfo {author} {\bibfnamefont {M.~A.}\
  \bibnamefont {Wolfe}}, \bibinfo {author} {\bibfnamefont {D.~E.}\ \bibnamefont
  {Savage}}, \bibinfo {author} {\bibfnamefont {M.~G.}\ \bibnamefont {Lagally}},
  \bibinfo {author} {\bibfnamefont {S.~N.}\ \bibnamefont {Coppersmith}},
  \bibinfo {author} {\bibfnamefont {M.}~\bibnamefont {Friesen}}, \bibinfo
  {author} {\bibfnamefont {R.}~\bibnamefont {Joynt}},\ and\ \bibinfo {author}
  {\bibfnamefont {M.~A.}\ \bibnamefont {Eriksson}},\ }\href
  {https://doi.org/10.48550/ARXIV.2112.09765} {\bibinfo {title} {Sige quantum
  wells with oscillating ge concentrations for quantum dot qubits}} (\bibinfo
  {year} {2021}{\natexlab{b}}),\ \Eprint {https://arxiv.org/abs/2112.09765}
  {arXiv:2112.09765} \BibitemShut {NoStop}%
\bibitem [{\citenamefont {Gu\'ery-Odelin}\ \emph {et~al.}(2019)\citenamefont
  {Gu\'ery-Odelin}, \citenamefont {Ruschhaupt}, \citenamefont {Kiely},
  \citenamefont {Torrontegui}, \citenamefont {Mart\'{\i}nez-Garaot},\ and\
  \citenamefont {Muga}}]{ShortcutsToAdiabaticity}%
  \BibitemOpen
  \bibfield  {author} {\bibinfo {author} {\bibfnamefont {D.}~\bibnamefont
  {Gu\'ery-Odelin}}, \bibinfo {author} {\bibfnamefont {A.}~\bibnamefont
  {Ruschhaupt}}, \bibinfo {author} {\bibfnamefont {A.}~\bibnamefont {Kiely}},
  \bibinfo {author} {\bibfnamefont {E.}~\bibnamefont {Torrontegui}}, \bibinfo
  {author} {\bibfnamefont {S.}~\bibnamefont {Mart\'{\i}nez-Garaot}},\ and\
  \bibinfo {author} {\bibfnamefont {J.~G.}\ \bibnamefont {Muga}},\ }\bibfield
  {title} {\bibinfo {title} {Shortcuts to adiabaticity: Concepts, methods, and
  applications},\ }\href {https://doi.org/10.1103/RevModPhys.91.045001}
  {\bibfield  {journal} {\bibinfo  {journal} {Rev. Mod. Phys.}\ }\textbf
  {\bibinfo {volume} {91}},\ \bibinfo {pages} {045001} (\bibinfo {year}
  {2019})}\BibitemShut {NoStop}%
\bibitem [{\citenamefont {Rabi}(1937)}]{RabiOriginal}%
  \BibitemOpen
  \bibfield  {author} {\bibinfo {author} {\bibfnamefont {I.~I.}\ \bibnamefont
  {Rabi}},\ }\bibfield  {title} {\bibinfo {title} {Space quantization in a
  gyrating magnetic field},\ }\href {https://doi.org/10.1103/PhysRev.51.652}
  {\bibfield  {journal} {\bibinfo  {journal} {Phys. Rev.}\ }\textbf {\bibinfo
  {volume} {51}},\ \bibinfo {pages} {652} (\bibinfo {year} {1937})}\BibitemShut
  {NoStop}%
\bibitem [{\citenamefont {Joecker}\ \emph {et~al.}(2019)\citenamefont
  {Joecker}, \citenamefont {Cerfontaine}, \citenamefont {Haupt}, \citenamefont
  {Schreiber}, \citenamefont {Kardyna\l{}},\ and\ \citenamefont
  {Bluhm}}]{RabiBenjamin}%
  \BibitemOpen
  \bibfield  {author} {\bibinfo {author} {\bibfnamefont {B.}~\bibnamefont
  {Joecker}}, \bibinfo {author} {\bibfnamefont {P.}~\bibnamefont
  {Cerfontaine}}, \bibinfo {author} {\bibfnamefont {F.}~\bibnamefont {Haupt}},
  \bibinfo {author} {\bibfnamefont {L.~R.}\ \bibnamefont {Schreiber}}, \bibinfo
  {author} {\bibfnamefont {B.~E.}\ \bibnamefont {Kardyna\l{}}},\ and\ \bibinfo
  {author} {\bibfnamefont {H.}~\bibnamefont {Bluhm}},\ }\bibfield  {title}
  {\bibinfo {title} {Transfer of a quantum state from a photonic qubit to a
  gate-defined quantum dot},\ }\href
  {https://doi.org/10.1103/PhysRevB.99.205415} {\bibfield  {journal} {\bibinfo
  {journal} {Phys. Rev. B}\ }\textbf {\bibinfo {volume} {99}},\ \bibinfo
  {pages} {205415} (\bibinfo {year} {2019})}\BibitemShut {NoStop}%
\end{thebibliography}%

\end{document}